\documentclass{aa}
\usepackage[varg]{txfonts}
\usepackage{graphicx}
\usepackage{natbib}
\usepackage{epstopdf}
\usepackage{units}
\usepackage[utf8]{inputenc}
\usepackage{hyperref}
\DeclareGraphicsExtensions{.pdf,.eps,.png,.jpg,.mps} 
\bibpunct{(}{)}{;}{a}{}{,}

\begin{document}
\title{Determination of fundamental asteroseismic parameters using the Hilbert transform}

\author{René Kiefer\inst{1}
\and Ariane Schad\inst{1}
\and Wiebke Herzberg\inst{1}
\and Markus Roth\inst{1}
}
\institute{Kiepenheuer-Institut für Sonnenphysik, Schöneckstraße 6, 79104 Freiburg, Germany}
\date{Received 5 December, 2014/ Accepted 25 March, 2015}
\titlerunning{Determination of asteroseismic parameters using the Hilbert transform}
\authorrunning{René Kiefer et al.}

\abstract{Solar-like oscillations exhibit a regular pattern of frequencies. This pattern is dominated by the small and large frequency separations between modes. The accurate determination of these parameters is of great interest, because they give information about e.g. the evolutionary state and the mass of a star.}
{We want to develop a robust method to determine the large and small frequency separations for time series with low signal-to-noise ratio. For this purpose, we analyse a time series of the Sun from the GOLF instrument aboard SOHO and a time series of the star KIC~5184732 from the NASA \textit{Kepler} satellite by employing a combination of Fourier and Hilbert transform.}
{We use the analytic signal of filtered stellar oscillation time series to compute the signal envelope. Spectral analysis of the signal envelope then reveals frequency differences of dominant modes in the periodogram of the stellar time series.}
{With the described method the large frequency separation $\Delta\nu$ can be extracted from the envelope spectrum even for data of poor signal-to-noise ratio. A modification of the method allows for an overview of the regularities in the periodogram of the time series.}
{}
\maketitle

\section{Introduction}\label{sec:1}
The space-borne \textit{Kepler} and \textit{CoRoT} missions have provided us with photometric data of exquisite quality for asteroseismic studies. The identification of solar-like oscillations in the power spectrum and the extraction of fundamental asteroseismic parameters is of great importance. However, this problem is not always straightforward, especially if the intrinsic properties of a star lead to a low signal-to-noise ratio of acoustic modes in the power spectrum.

Asteroseismic parameters, like the frequency of maximum power $\nu_{\text{max}}$ and the large frequency separation $\Delta\nu$, are needed to estimate the mass and radius of the investigated stars with the use of scaling relations \citep[e.g.][]{2013A&A...550A.126M}. These scaling relations yield very good results for solar-like stars \citep[e.g.][]{2012ApJ...749..152M}. In the literature there are many different approaches to detect the required parameters, e.g. \cite{2012ApJ...749..152M, 2011arXiv1104.0631V, 2009A&A...508..877M, 2010A&A...511A..46M,2009CoAst.160...74H, 2009A&A...508..877M}. A comparison of complementary techniques that are used to extract global asteroseismic parameters was performed by e.g. \cite{2011MNRAS.415.3539V} and \cite{2011A&A...525A.131H}. All of these methods either use the fitting of individually resolved modes or the autocorrelation of the power spectrum. 

The mean large frequency separation $\Delta\nu$ between consecutive radial orders dominates the comb-like pattern, which is characteristic for solar-like oscillations.  According to \cite{1980ApJS...43..469T} the eigenfrequency of radial order $n$ and harmonic degree $l$ is given in this asymptotic theory by 
\begin{align}
\nu_{nl}=\left[n+l/2+\epsilon\right]\Delta\nu-l(l+1)D_0\ .
\end{align}
Here $\Delta\nu$ is the large frequency separation, $D_0$ is a second order term that determines the small separation $\delta\nu$, and $\epsilon$ is a constant term. If the modes are resolved well in the periodogram, a direct method to estimate $\Delta\nu$ is to fit individual modes and to calculate the large frequency separation from the fit parameters. This way the observed frequency dependence of $\Delta\nu$, which arises in second order calculations of the asymptotic theory \citep{2013A&A...550A.126M}, can be measured. \cite{2009A&A...508..877M} presented an autocorrelation method that is capable of tracking the variation of $\Delta\nu$ with frequency without fitting individual modes. The method we present here can be used for detection of the mean $\Delta\nu$ and is adaptable for the mapping of $\Delta\nu(\nu)$ and the frequency dependence of other regularities in the periodogram.

In Sect.~\ref{sec:2} we describe a novel method to detect regularities in the periodogram. In Sect.~\ref{sec:3} the method is applied to stellar time series, i.e. Sun-as-a-star data from the GOLF instrument and data for the solar-like \textit{Kepler} star KIC~5184732. An investigation of the robustness of the method for low signal-to-noise ratio power spectra and a comparison to the autocorrelation of the periodogram are presented in Sect.~\ref{sec:3.4} and Sect.~\ref{sec:3.5}, respectively. We provide a discussion of the results and a conclusion in Sect.~\ref{sec:4}.  

\section{The envelope spectrum}\label{sec:2}
From acoustics, the beat phenomenon is very well known. The superposition of two oscillating signals with different frequencies 
results in a fast oscillating component with a slowly modulated amplitude. For the case of two cosines of equal amplitude one can write
\begin{align}
 \cos\left(\omega_1 t\right) + \cos\left(\omega_2 t\right) &= A\left(t\right) \cos\left(\frac{1}{2}\left(\omega_1 + \omega_2\right)t\right)\,,\\
A\left(t\right) &= 2 \cos\left(\frac{1}{2}\left(\omega_1 - \omega_2\right)t\right)\,,
\end{align}
where $A(t)$ is the slowly varying amplitude of the signal which oscillates with frequency
\begin{align}
 \Omega =\frac{1}{2}\left(\omega_1 - \omega_2 \right)\,.\label{eq:beatfreq}
\end{align}

More generally, for a multi-component signal with $n$ oscillating components, where each has an analytic representation, one obtains a series of products over cosines of pairwise sums and differences of the involved frequencies $\omega_k$ with $k=1,\dots,n$. The terms $\cos(\omega_{i}-\omega_{j})$ with $i\not= j$ contribute to a low-frequency modulation of the signal amplitude and forms the signal envelope. Utilising this envelope, it is possible to measure frequency differences of modes present in a time series, as described in the following. First, the periodogram of the investigated time series is computed. All phase information is neglected this way. Next, the square root of the whole periodogram or a filtered range of interest is transformed back to the temporal domain by an inverse Fourier transform, thereby resulting in a new time series $x(t)$. Taking the square root of the periodogram ensures correct physical dimensions. The time series $x(t)$ is a superposition of all the modes in the inversely transformed range of the periodogram. The modes present in $x(t)$ all have the same phase lag $\phi_{0}=0$. 

The analytic signal $x_a\left(t\right)$ of a real valued signal $x\left(t\right)$ is defined by
\begin{align}
x_a(t) &= x(t) + i\, \mathrm{H}\left[x\right](t)\, ,\\
\mathrm{H}\left[x\right](t) &= x(t) \ast \frac{1}{\pi t} = \frac{1}{\pi} \int\limits_{-\infty}^\infty \frac{x(\tau)}{t-\tau} \mathrm{d}\tau\, ,
\end{align}
where $\mathrm{H}\left[x\right](t)$ is the Hilbert transform. The $\ast$ operator indicates convolution. By taking the modulus of $x_a(t)$, the envelope of $x(t)$ is computed. 

\begin{figure}[ht]
\centering{
\includegraphics[angle=270,width=0.49\textwidth]{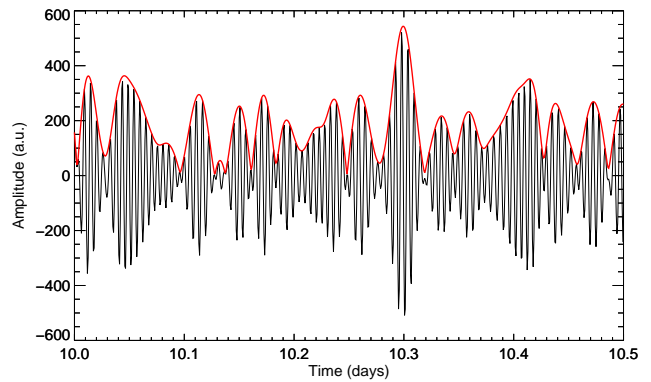}
\caption{Segment of the time series $x(t)$ for KIC 5184732 (black) and the corresponding signal envelope (red). The frequency range in which \cite{2012ApJ...749..152M} reported observed p-mode oscillations was chosen as boundaries for the frequency filter (\unit[1.4--2.7]{mHz}).}
\label{fig:1}}
\end{figure}

In the spectrum of $\text{abs}(x_a(t))$, henceforth called the envelope spectrum, the frequency difference of regularly spaced modes in the spectrum of the original time series appears as strong peaks at the position defined by twice the value given by Eq.~\ref{eq:beatfreq}. The reason for this is that Eq.~\ref{eq:beatfreq} describes the frequency of the whole modulated amplitude $A(t)$, but the envelope considers only the positive side of $x(t)$ (compare Fig.~\ref{fig:1}). The corresponding peak in the envelope spectrum is more pronounced the more often a given regular spacing appears in the periodogram of the original time series. This way even regularly spaced modes with low amplitude produce a peak in the envelope spectrum.

In Fig.~\ref{fig:1} a segment of $x(t)$ with a length of half a day day and its envelope $\text{abs}(x_a(t))$ are shown for the Kepler star KIC~5184732. The periodogram was filtered for the frequency range of p-modes \citep{2012ApJ...749..152M}.

\section{Application}\label{sec:3}
\subsection{Application to GOLF data}\label{sec:3.1}
We use Sun-as-a-star data, i.e. integrated light data, from the GOLF (Global oscillations a low frequency) instrument, which is aboard the SOHO satellite \citep{2005A&A...442..385G} for our analyses of the solar frequency separations. The temporal cadence of GOLF data is \unit[20]{s}. We use a year-long segment of the full GOLF time series covering July 2007 to July 2008. 

For the GOLF data we apply a fast Fourier transform (FFT) to calculate the periodogram. To increase the signal-to-noise ratio in the final envelope spectrum, the data are filtered by multiplication of the periodogram with a \textit{Tukey} window with $\alpha = 0.9$ \citep{Harris}, rather than by a boxcar window. The analysis is carried out in the spectral range of \unit[1.7--3.5]{mHz}, which approximately corresponds to the range of strong p-modes in the periodogram. The parameters of width and location of this band-pass filter are chosen in such a way that all frequencies but the band of interest are set to zero.

\begin{figure}[ht]
\centering{
\includegraphics[angle=270,width=0.49\textwidth]{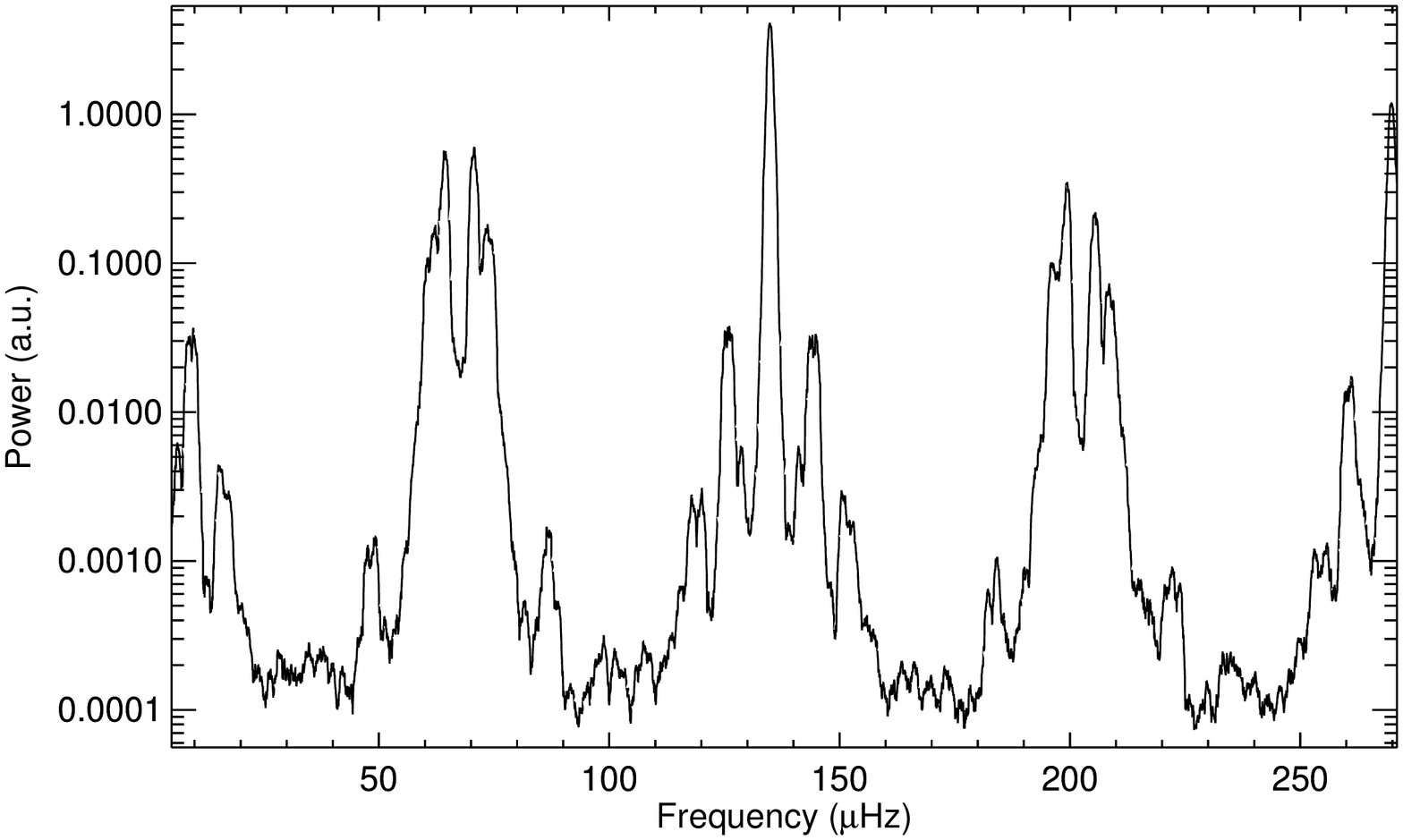}\\
\includegraphics[angle=270,width=0.49\textwidth]{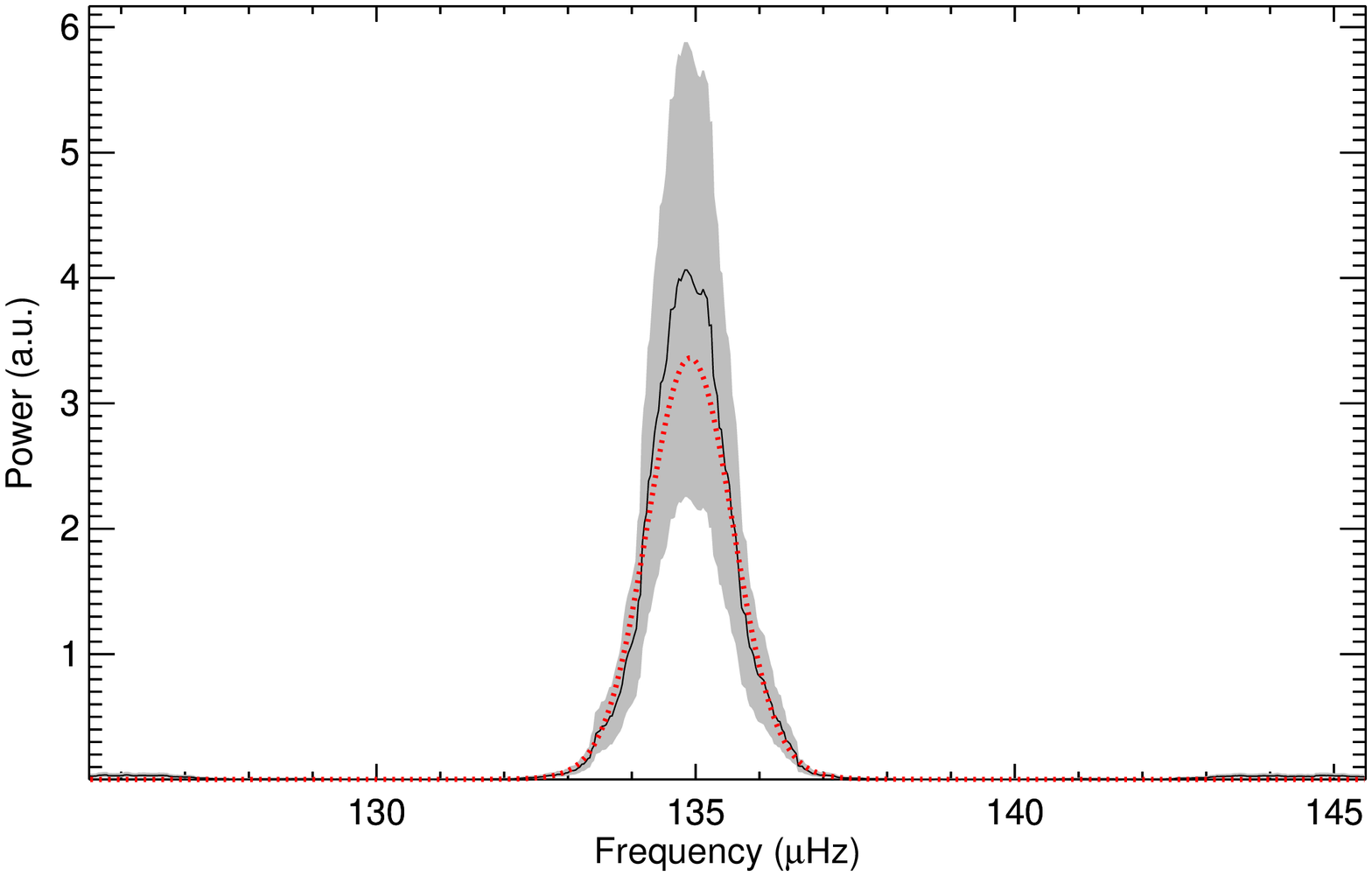}
\caption{\textbf{Top panel:} Semi-logarithmic plot of the low frequency part of the $\unit[1]{\mu Hz}$ boxcar smoothed envelope spectrum for GOLF data. The periodogram was filtered for the frequency range of strong p-modes between \unit[1.7--3.5]{mHz}. \textbf{Bottom panel:} Fit of Voigt profile (dashed red line) to the peak of the large frequency separation in the envelope spectrum with $2$-$\sigma$ confidence intervals in grey colour.}
\label{fig:2}}
\end{figure}

As we discussed in Sect.~\ref{sec:2}, the envelope spectrum features peaks at frequency values that correspond to differences between frequencies $|\nu_{1}-\nu_{2}|$ of modes with sufficiently high amplitude in the original periodogram or if the filtered frequency band contains many instances of equidistant spacing. The envelope spectrum for the analysed GOLF time series is displayed in Fig.~\ref{fig:2}. The strong peak at $\approx\,$\unit[135]{$\mu$Hz} corresponds to the large frequency separation $\Delta\nu$. The peak structure at $\approx\,$\unit[70]{$\mu$Hz} is due to the separation of modes, which are listed in the columns 5--8 of Table~\ref{table:2} (a detailed discussion follows below). Since many radial orders are within the chosen frequency range, overtones can be identified at multiples of the discussed frequency separations. The abscissa is truncated at a frequency of \unit[5]{$\mu$Hz} because frequency differences of this regime occur within the linewidth of p-mode peaks in the periodogram and therefore have large amplitude in the envelope spectrum. At the upper end the abscissa is truncated at $2\cdot\Delta\nu$, so only the first overtone for $\Delta n=2$ is visible.

\begin{figure}[ht]
\centering{
\includegraphics[angle=270,width=0.49\textwidth]{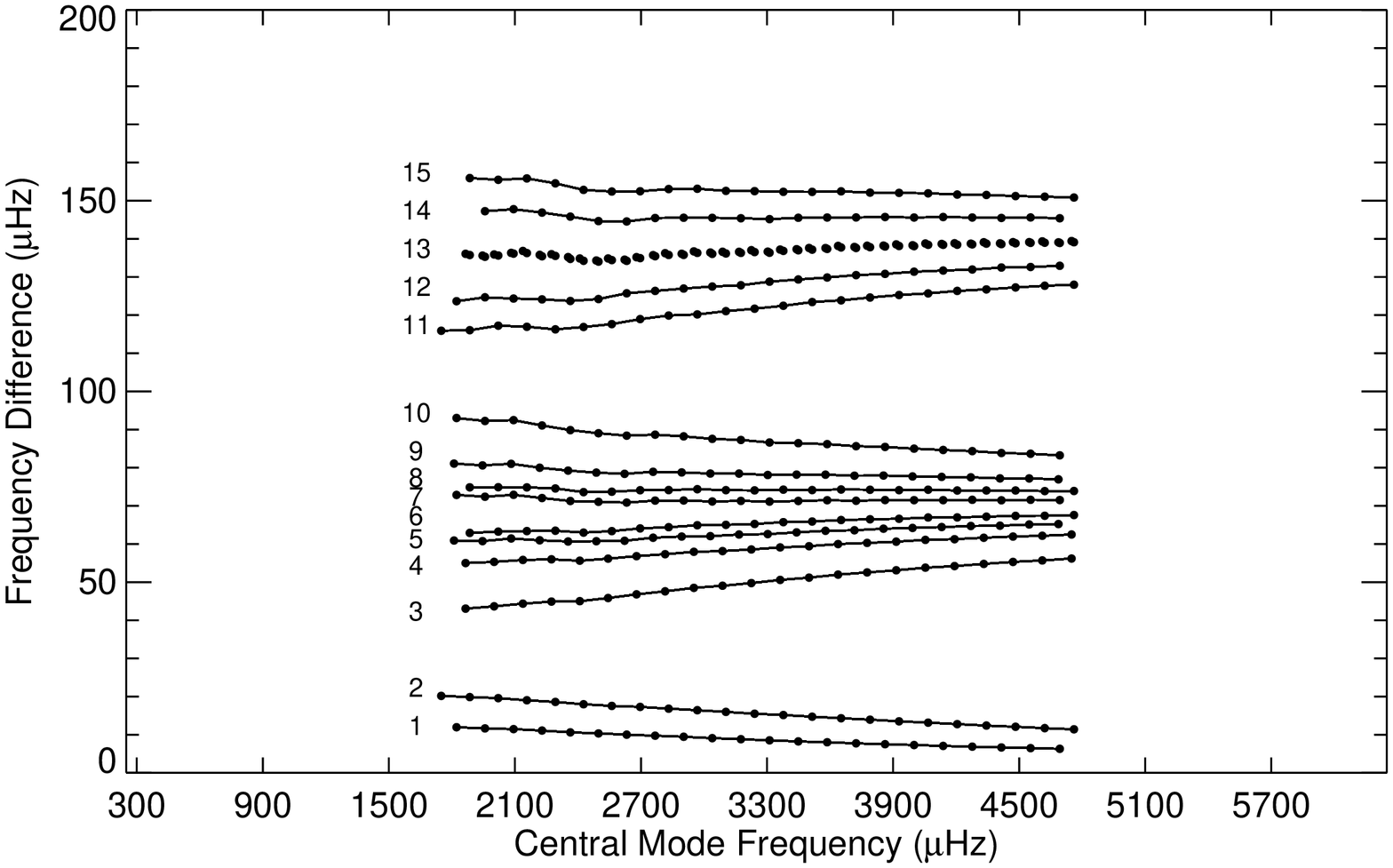}
\includegraphics[angle=270,width=0.49\textwidth]{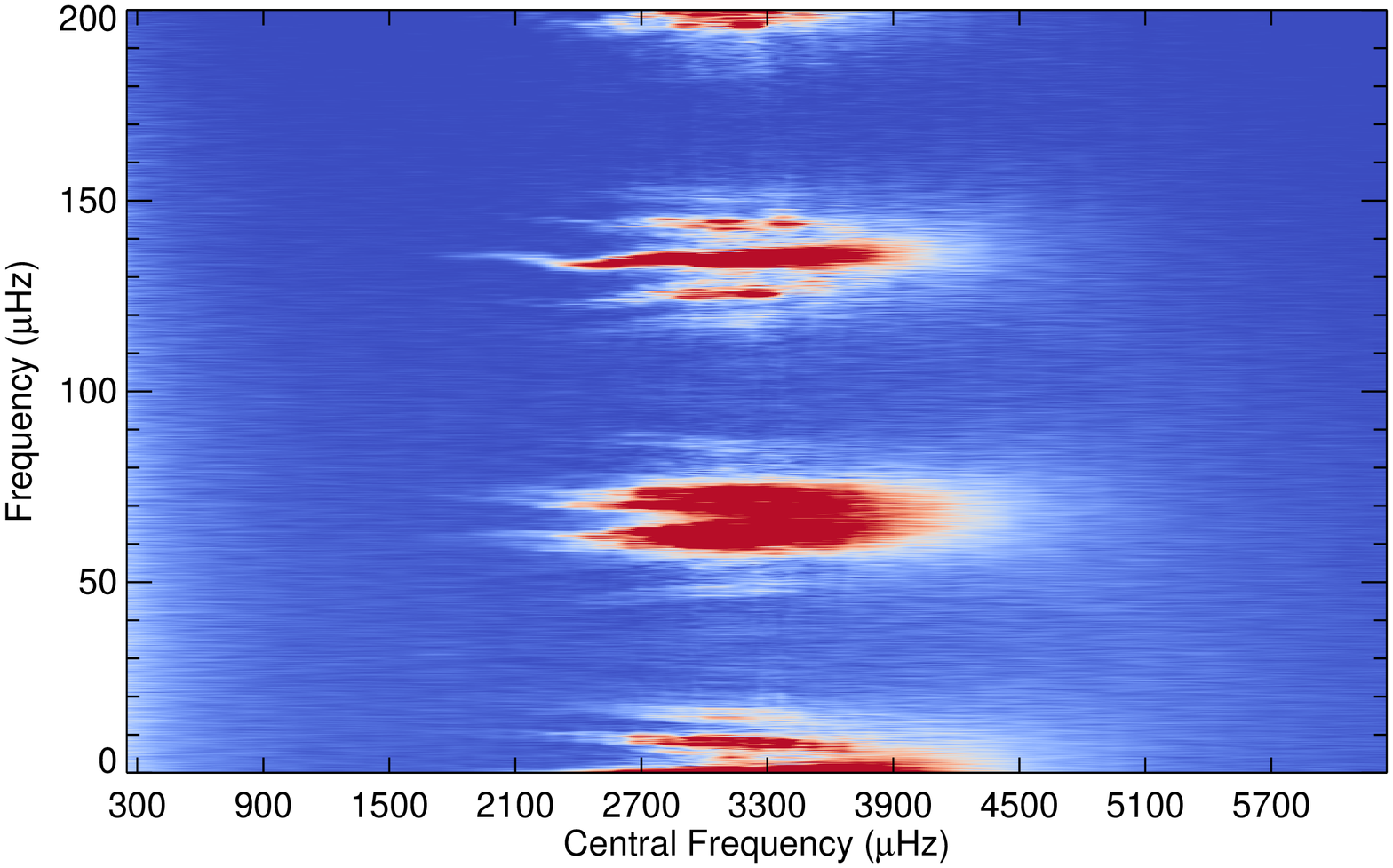}
\caption{\textbf{Top panel:} Frequency differences between modes as a function of central mode frequency. Only modes with harmonic degree $l\le 3$ and \unit[1750]{$\mu$Hz}\,$\le\nu\le$\,\unit[4750]{$\mu$Hz} are considered. The modes of which the frequencies are subtracted, resulting in the numbered bands, are listed in Table~\ref{table:2}. Mode frequencies from \cite{1996Sci...272.1286C}. \textbf{Bottom panel:} Map of the solar envelope spectrum as a function of central frequency of the used spectral filter which has a width of \unit[500]{$\mu$Hz}. A vertical cut at any central frequency value through this diagram is the respective envelope spectrum. Boxcar smoothed in y direction over $\unit[0.2]{\mu Hz}$. Red colour represents a high amplitude in the envelope spectrum, blue a low amplitude.}
\label{fig:3}}
\end{figure} 

\begin{table} 
\caption{List of modes frequencies used to calculate differences.}           
\label{table:2}   
\centering                                   
\begin{tabular}{c  c  c }      
\hline\hline
Band &  Modes  & Name\\ 
\hline
1 	& $\nu_{n,0}-\nu_{n-1,2}$ & $\delta\nu_{02}$ \\
2 	& $\nu_{n,1}-\nu_{n-1,3}$ & $\delta\nu_{13}$ \\
3 	& $\nu_{n,3}-\nu_{n+1,0}$ & \\
4 	& $\nu_{n,3}-\nu_{n,2}$ & \\
5 	& $\nu_{n,2}-\nu_{n,1}$ & \\
6 	& $\nu_{n,1}-\nu_{n,0}$ & \\
7 	& $\nu_{n,0}-\nu_{n-1,1}$ & \\
8 	& $\nu_{n,1}-\nu_{n-1,2}$ & \\
9 	& $\nu_{n,2}-\nu_{n-1,3}$ & \\
10 	& $\nu_{n,0}-\nu_{n-2,3}$ & \\
11	& $\nu_{n,1}-\nu_{n,3}$ & $\Delta\nu-\delta\nu_{13}$ \\
12 	& $\nu_{n,0}-\nu_{n,2}$ & $\Delta\nu-\delta\nu_{02}$ \\
13 	& $\nu_{n,l}-\nu_{n-1,l}$ & $\Delta\nu$ \\
14	& $\nu_{n,0}-\nu_{n-2,2}$ & $\Delta\nu+\delta\nu_{02}$ \\
15	& $\nu_{n,1}-\nu_{n-2,3}$ & $\Delta\nu+\delta\nu_{13}$ \\
\hline                                           
\end{tabular}
\tablefoot{First column: Number of band in the top panel of Fig.~\ref{fig:3}. Second column: Mode frequencies used to calculate the differences. Third column: Common notation of important frequency separations.}
\end{table}

A scan for a possible frequency dependence of regularities in the periodogram, e.g. $\Delta\nu(\nu)$, or locating the spectral range of regularities, can be done by choosing a narrow spectral filter and incrementally shifting the filter through the periodogram. For an investigation of the variation of the solar frequency separations we set the width of the spectral filter to \unit[500]{$\mu$Hz} and the shift increment to \unit[10]{$\mu$Hz}. The bottom panel of Fig.~\ref{fig:3} shows the result, representing a map of the regularities in the periodogram as a function of central filter frequency. This map is produced without any previous knowledge of the frequency separations in the periodogram. Therefore, its calculation is less dependent on critical parameters as the common echelle diagram, which needs an accurate estimate of $\Delta\nu$ to be of diagnostic value.

The top panel of Fig.\ref{fig:3} depicts all differences of modes with harmonic degree $l\le 3$ and \unit[1750]{$\mu$Hz}\,$\le\nu\le$\,\unit[4750]{$\mu$Hz}. Mode frequencies were taken from \cite{1996Sci...272.1286C}. Each numbered band in the top panel of Fig.~\ref{fig:3} corresponds to the frequency difference of the modes given in the respective row in Table~\ref{table:2}. The mode linewidth of the peaks in the filtered frequency range correspond to the red structure at the lowest frequency values. As expected~\citep{1999A&A...351..582H, 2000ApJ...543..472K}, the mode linewidth increases with higher frequency which results in a broadening of all features in this diagram towards higher central filter frequency values. The isolated feature around frequency values of $\approx\,$\unit[10]{$\mu$Hz} corresponds to the small frequency separation $\delta\nu_{02}$. The feature exhibits a negative slope, which corresponds to a decrease of $\delta\nu_{02}$ with higher frequency. The second band results from the small frequency separation $\delta\nu_{13}$. Like all features in this diagram for which one of the mode frequencies is of harmonic degree $l=3$ (bands 2--4, 9--11, 15) the amplitude is smaller than for bands that result from the difference of frequencies of modes of lower harmonic degree, since the amplitude of $l=3$ modes is considerably smaller in integrated light. 

At frequency values between \unit[50--80]{$\mu$Hz} there is a double peak structure which can be explained as follows: The double structure merges towards higher central frequency values due to the increasing linewidths and the decreasing frequency separations of the modes in the periodogram. The differences of modes with lower frequency difference $(\nu_{n,l} - \nu_{n,l-1})$ (bands 4--6) and higher frequency difference $(\nu_{n,l}-\nu_{n-1,l+1})$ (bands 7--9) cause the main peaks. If the frequency differences which belong to bands 5 and 6 as well as those of bands 7 and 8 have approximately the same value, these two double bands are separated by $\delta\nu_{02}$ from each other. 

The large amplitude structure around $\approx\,$\unit[135]{$\mu$Hz} is due to $\Delta\nu$
and resembles the course of the theoretical values for the frequency differences which are shown in the top panel of Fig.~\ref{fig:3}. The lower and upper sidelobes of this structure correspond to frequency differences between modes with $\left(\nu_{n,0} - \nu_{n,2}\right)$ (band 12) and $\left(\nu_{n,0} - \nu_{n-2,2}\right)$ (band 14), respectively. Thus, these two bands are separated from the band of the large frequency separation by $\delta\nu_{02}$. Two small amplitude bands (11 and 15) can be spotted at a $\Delta\nu\mp\delta\nu_{13}$.

\subsection{Parameter extraction}
Frequency separations are estimated from the envelope spectrum in the following manner. All modes of same degree and consecutive radial order are separated by $\Delta\nu$, e.g. two modes of degree $l=0$ which differ in radial order by $\Delta n=1$. Therefore, the peak at $\nu\approx\unit[135]{\mu Hz}$ present in Fig.~\ref{fig:2} accounts for a large number of radial orders. Furthermore, there are peaks at multiples of $\Delta\nu$ present in the envelope spectrum. These peaks account for modes of same degree but a radial order which differs by $\Delta n=2,3,\dots$. In Fig.~\ref{fig:2}, only the first overtone for $\Delta n=2$ is visible, since the frequency axis is truncated at $2\cdot\Delta\nu$. 

In order to fit for the peak frequencies in the envelope spectrum we select an area around the peaks of interest with a width of \unit[15]{$\mu$Hz}. Confidence bands on the boxcar smoothed envelope spectrum are computed according to \cite{Priestley}:
\begin{align}
\left[\frac{nP(\nu)}{b_{n}(\alpha)},\frac{nP(\nu)}{a_{n}(\alpha)}\right]\,,
\end{align}
where $n$ is the number of bins the periodogram $P(\nu)$ is smoothed over, and $a_{n}(\alpha)$ and $b_{n}(\alpha)$ denote the lower and upper $100(\alpha/2)\%$ quantiles of the $\chi_{n}^2$ distribution, respectively. 
The peaks are fitted with a Voigt profile. 

We tested three common peak profiles (Gaussian, Lorentzian, Voigt) on the envelope spectrum. Although the Voigt profile is not matching the height of the peaks perfectly in some cases (see e.g. bottom panel of Fig.~\ref{fig:2}), it best fits the tails of the peaks in the envelope spectrum and provided the lowest $\chi^2_{\textrm{red}}$. Additionally, the Voigt profile gives a more conservative error bar on the peak center due to its higher half-width-at-half-maximum compared to the Gaussian and Lorentzian profiles.

Fitting not only the first overtone but also as many overtones as possible reduces the error on the fit parameters by computing the weighted mean:
\begin{align}
\Delta\nu &= \frac{\sum\limits_{k=1}^{n} \frac{\Delta\nu_{k}}{{s_{\Delta\nu_{k}}}^2}}{\sum\limits_{k=1}^{n}\frac{1}{{s_{\Delta\nu_{k}}}^2}}\, ,\\
s_{\Delta\nu} &= \frac{1}{\sqrt{\sum\limits_{k=1}^{n}\frac{1}{{s_{\Delta\nu_{k}}}^2}}}\, ,
\end{align}
where n is the number of overtones identified in the envelope spectrum. The individual large frequency separations and their errors are calculated by 
\begin{align}
\Delta\nu_{k} &= \frac{\nu_{\textrm{fit}}}{\textrm{k}}\, ,\\
s_{\Delta\nu_{k}} &= \frac{s_{\nu_{\textrm{fit}}}}{\textrm{k}}\, ,
\end{align}
where $\nu_{\textrm{fit}}$ is the fitted value at the k'th overtone and $s_{\nu_{\textrm{fit}}}$ is the half-width-at-half-maximum (HWHM) of this fit.

By applying this to the solar envelope spectrum, which is shown in Fig.~\ref{fig:2}, we obtain 
\begin{align}
 \Delta\nu_{\odot} = \unit[134.92\pm 0.06]{\mu Hz}\, ,
\end{align}
for the mean solar large frequency separation. The values found for $\Delta\nu_{\odot}$ in the literature are not fixed to one value. For example, \cite{2010A&A...522A...1K} use $\Delta\nu_{\odot}=\unit[134.88 \pm 0.04]{\mu Hz}$, while \cite{2013A&A...550A.126M} set $\Delta\nu_{\odot}=\unit[135.5]{\mu Hz}$ for their (uncalibrated) solar value.

\begin{figure*}[ht]
\begin{minipage}[t]{0.49\textwidth}
\includegraphics[angle=270,width=1.05\textwidth]{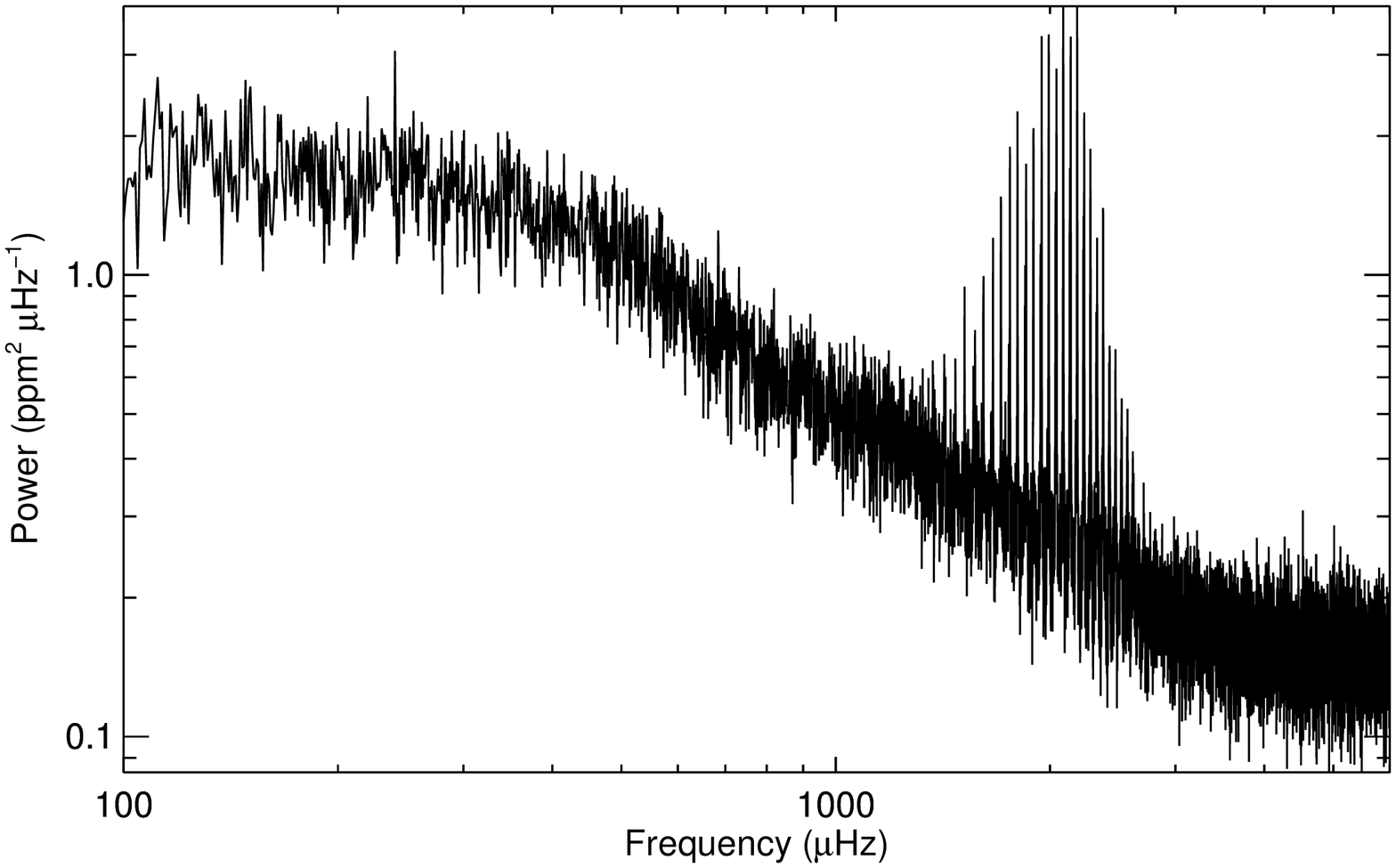}\vspace{-1.em}\\\vspace{-1.em}
\includegraphics[angle=270,width=1.05\textwidth]{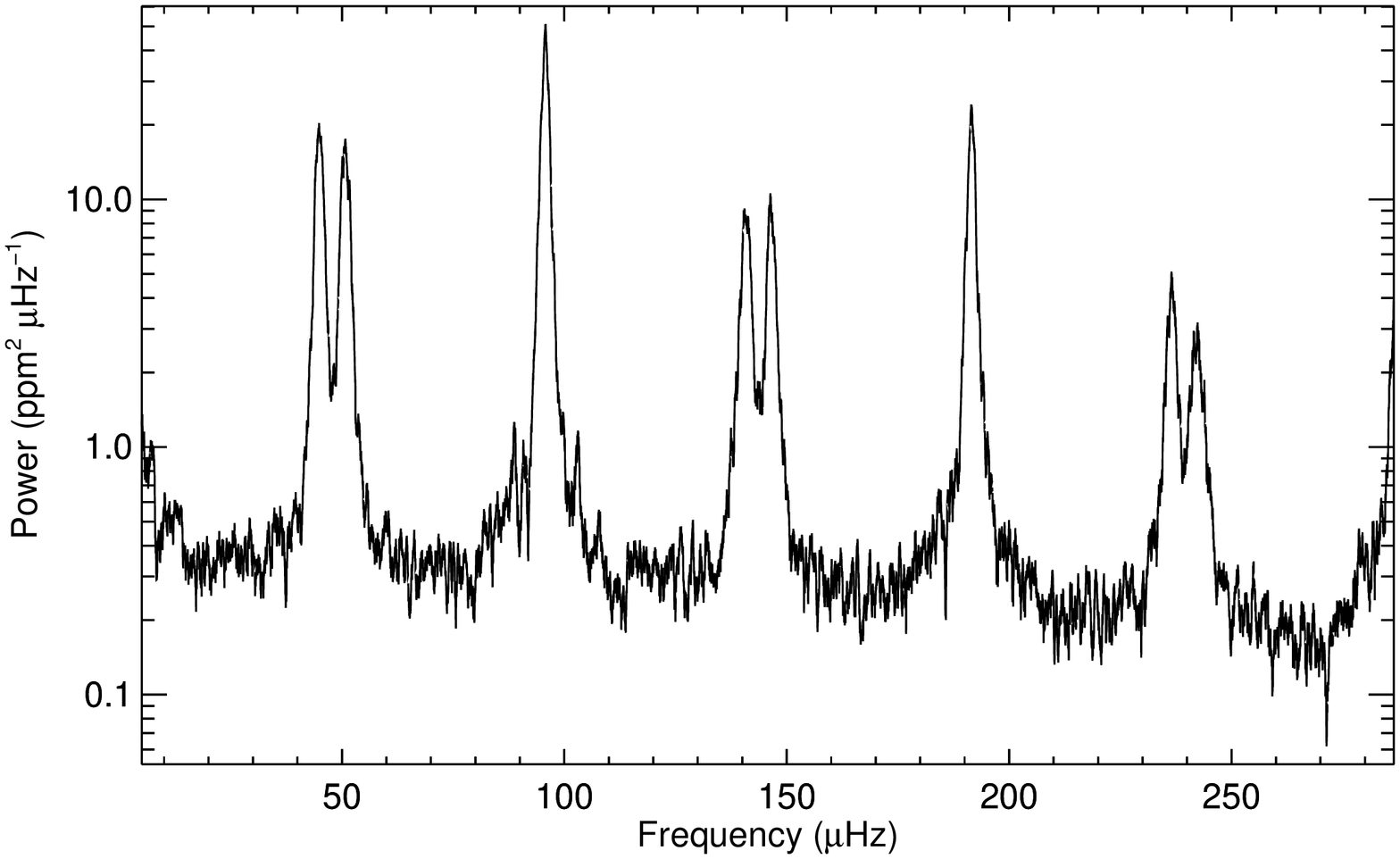}\\
\end{minipage}\hfill
\begin{minipage}[t]{0.49\textwidth}
\includegraphics[angle=270,width=1.05\textwidth]{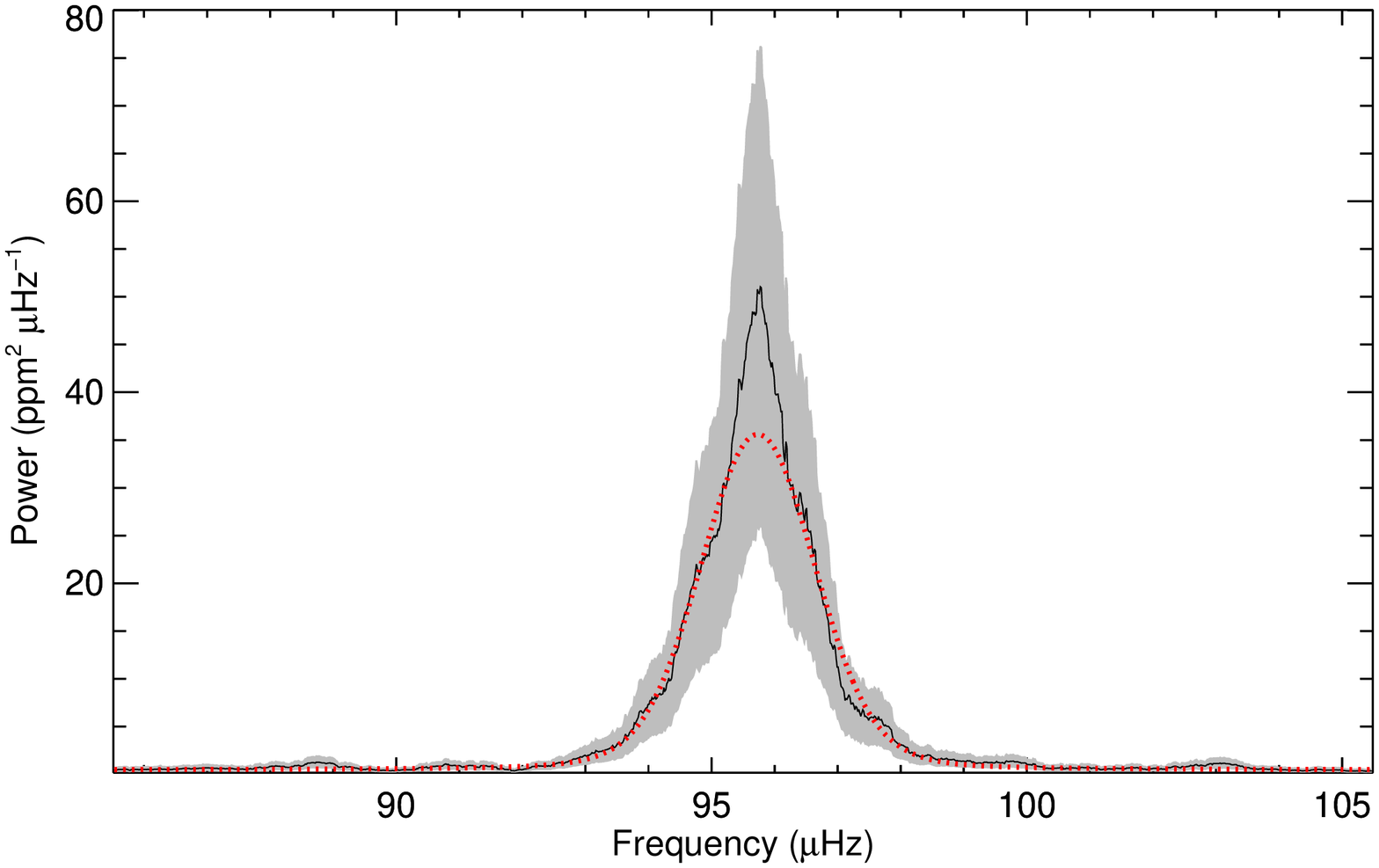}\vspace{-1.em}\\\vspace{-1.em}
\includegraphics[angle=270,width=1.05\textwidth]{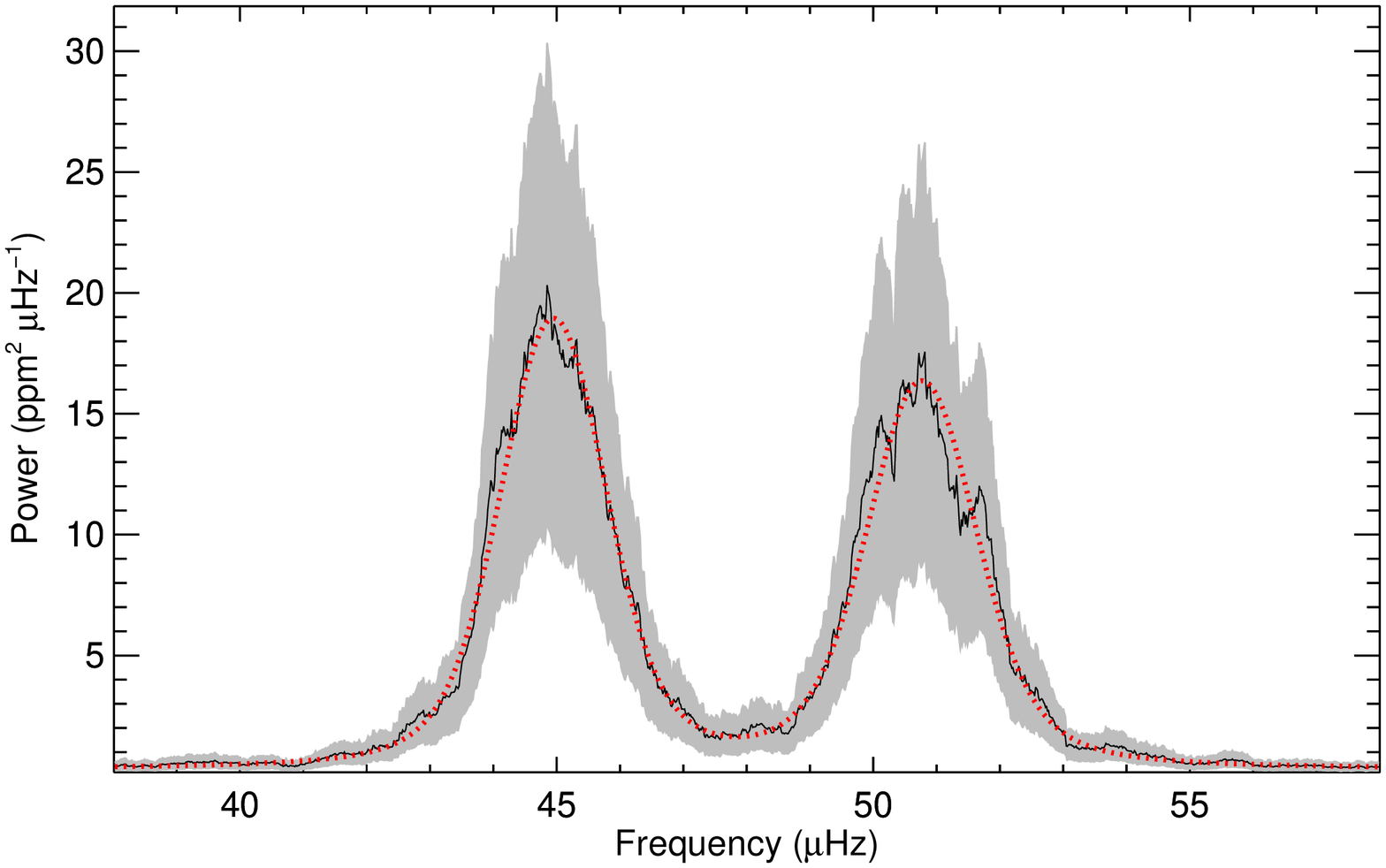}\\\vspace{1em}
\end{minipage}
\caption{\textbf{Top left panel: }Double-logarithmic plot of the spectrum for KIC~5184732. Boxcar smoothed over $\unit[0.5]{\mu Hz}$. \textbf{Bottom left panel:} Spectrum of the signal envelope shown in Fig.~\ref{fig:1} on a semi logarithmic scale. Peaks from dominant beat frequencies stand out against the background. Boxcar smoothed over $\unit[0.5]{\mu Hz}$. \textbf{Top right panel:} Segment of the envelope spectrum around the peak which is due to the large frequency separation (black) with $2$-$\sigma$ confidence intervals in grey colour. The Voigt fit to the peak is shown in red (dashed line). \textbf{Bottom right panel:} Segment of the envelope spectrum around the peaks from which the small separation can be computed (black) with $2$-$\sigma$ confidence intervals in grey colour. The Voigt fit to the peaks is shown in red (dashed line). }
\label{fig:4}
\end{figure*}

\subsection{Application to \textit{Kepler} data of KIC~5184732}\label{section:kic}
We apply our method to \textit{Kepler} data of KIC~5184732. The data cover the Kepler quarters Q7-15, i.e. approximately two years of observations. We use the short cadence data with a temporal cadence of \unit[58.85]{s}. Since \textit{Kepler} data are not perfectly equally spaced and the number of missing data points is not negligible, the periodogram is computed with the Lomb-Scargle method \citep{1976Ap&SS..39..447L, 1982ApJ...263..835S}. The rest of the analysis procedure is the same as described in the previous section, including the filter configuration. In the top left panel of Fig.~\ref{fig:4} the spectrum, i.e. the smoothed Lomb-Scargle-periodogram, of KIC~5184732 is shown. The p-mode region is clearly visible as an excess of power in the frequency range between \unit[1.4--2.7]{mHz}. This frequency range is chosen for the calculation of the envelope spectrum which is depicted in the bottom left panel of Fig.~\ref{fig:4}. The large frequency separation of KIC~5184732 can be estimated by fitting the peak at $\approx\,$\unit[95]{$\mu$Hz} and multiples of it. For \textit{Kepler} data, the number of overtones that can be fitted in the envelope spectrum varies from star to star. In the case of KIC~5184732 three overtones can be fitted in the envelope spectrum. The top right panel of Fig.~\ref{fig:4} shows a zoom into the region around the peak which corresponds to the large frequency separation. The Voigt fit is plotted in red. 

For the large frequency separation we obtain a value of
\begin{align}
\Delta\nu = \unit[95.7\pm 0.2]{\mu Hz}\, .
\end{align}
This is in accordance with the value of \cite{2012ApJ...749..152M}, who measure $\Delta\nu = 95.53\pm 0.26\,\mu$Hz. The small frequency separation can be calculated as the difference between the frequencies of the two peaks at $\approx\,$\unit[50]{$\mu$Hz}. A fit to these two peaks is depicted in the lower right panel of Fig.~\ref{fig:4}. For the small frequency separation, we obtain 
\begin{align}
\delta\nu_{02} = \unit[5.8\pm 1.0]{\mu Hz}\, .
\end{align}
In this case, an extraction of the small separation by fitting these two peaks is possible because the frequency separations of KIC~5184732 meet the criteria stated in Sect.~\ref{sec:3.1}.

\begin{figure}
\centering{
\includegraphics[angle=270,width=0.49\textwidth]{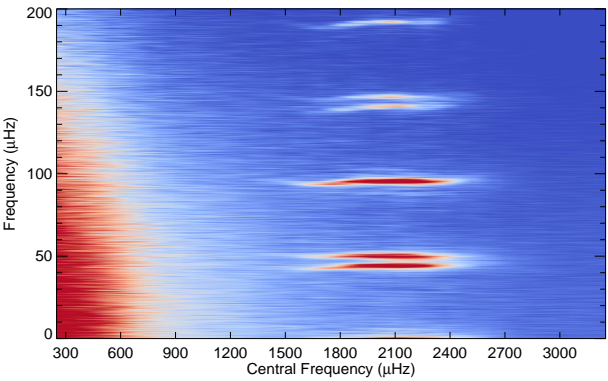}
\caption{Map of the envelope spectrum of KIC~5184732 as a function of central frequency of the used spectral filter which has a width of \unit[500]{$\mu$Hz}. A vertical cut at any central frequency value through this diagram is the respective envelope spectrum. Boxcar smoothed in y direction over $\unit[0.1]{\mu Hz}$. Red colour represents a high amplitude in the envelope spectrum, blue a low amplitude.}
\label{fig:5}}
\end{figure} 

\begin{figure*}[ht]
\begin{minipage}[t]{0.33\textwidth}
\includegraphics[angle=270,width=1.05\textwidth]{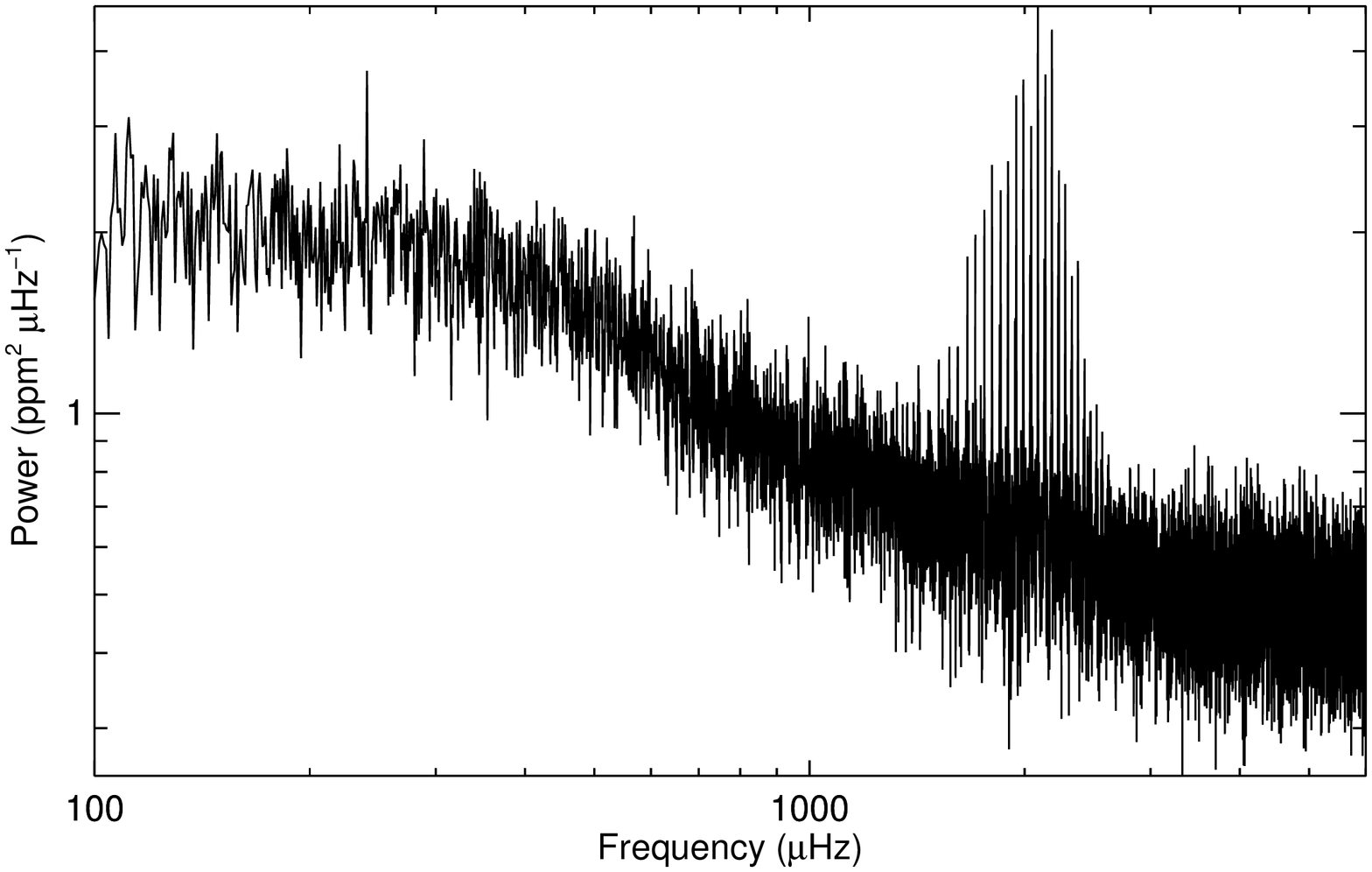}\vspace{-1.em}\\\vspace{-1.em}
\includegraphics[angle=270,width=1.05\textwidth]{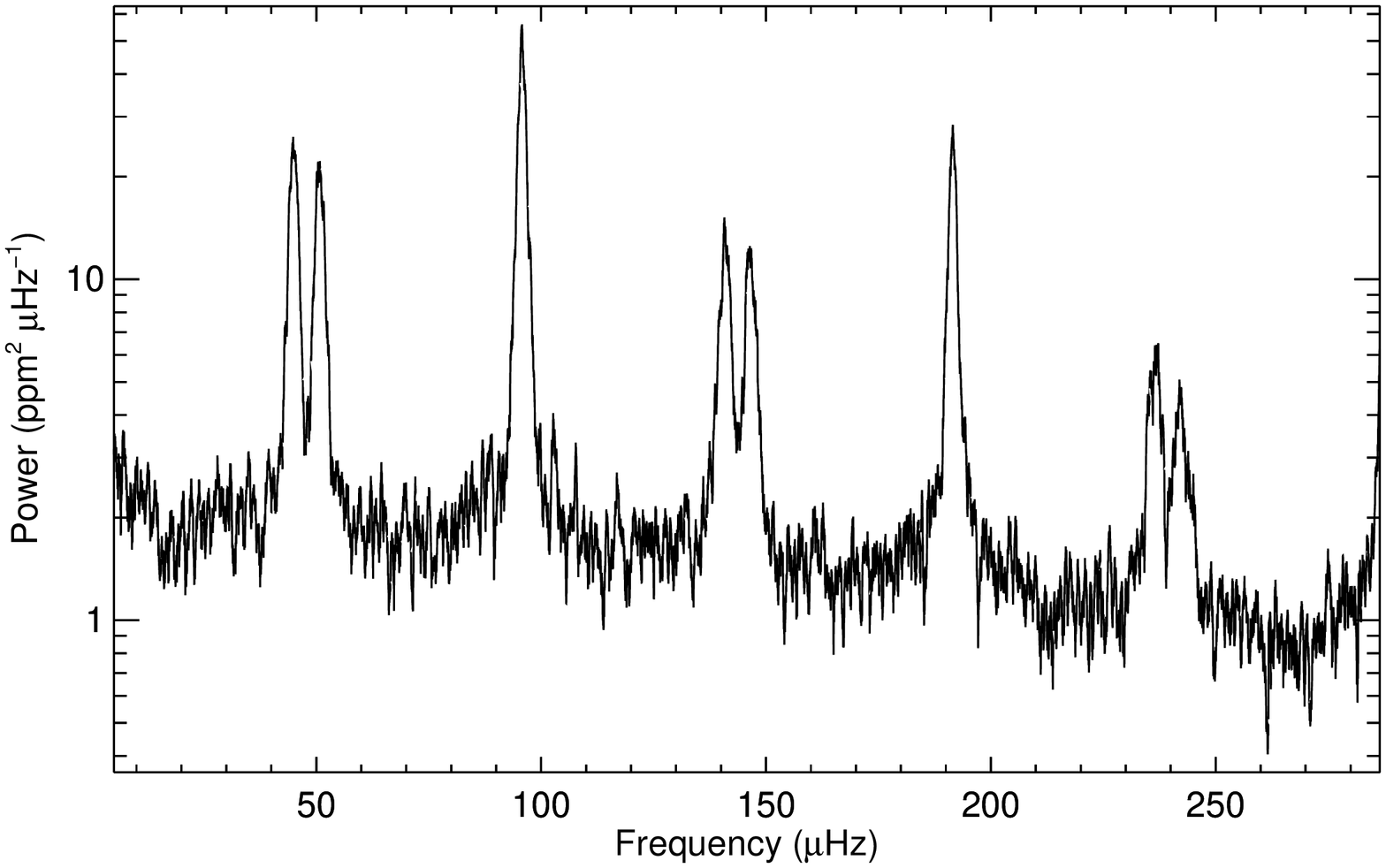}\\
\includegraphics[angle=270,width=1.05\textwidth]{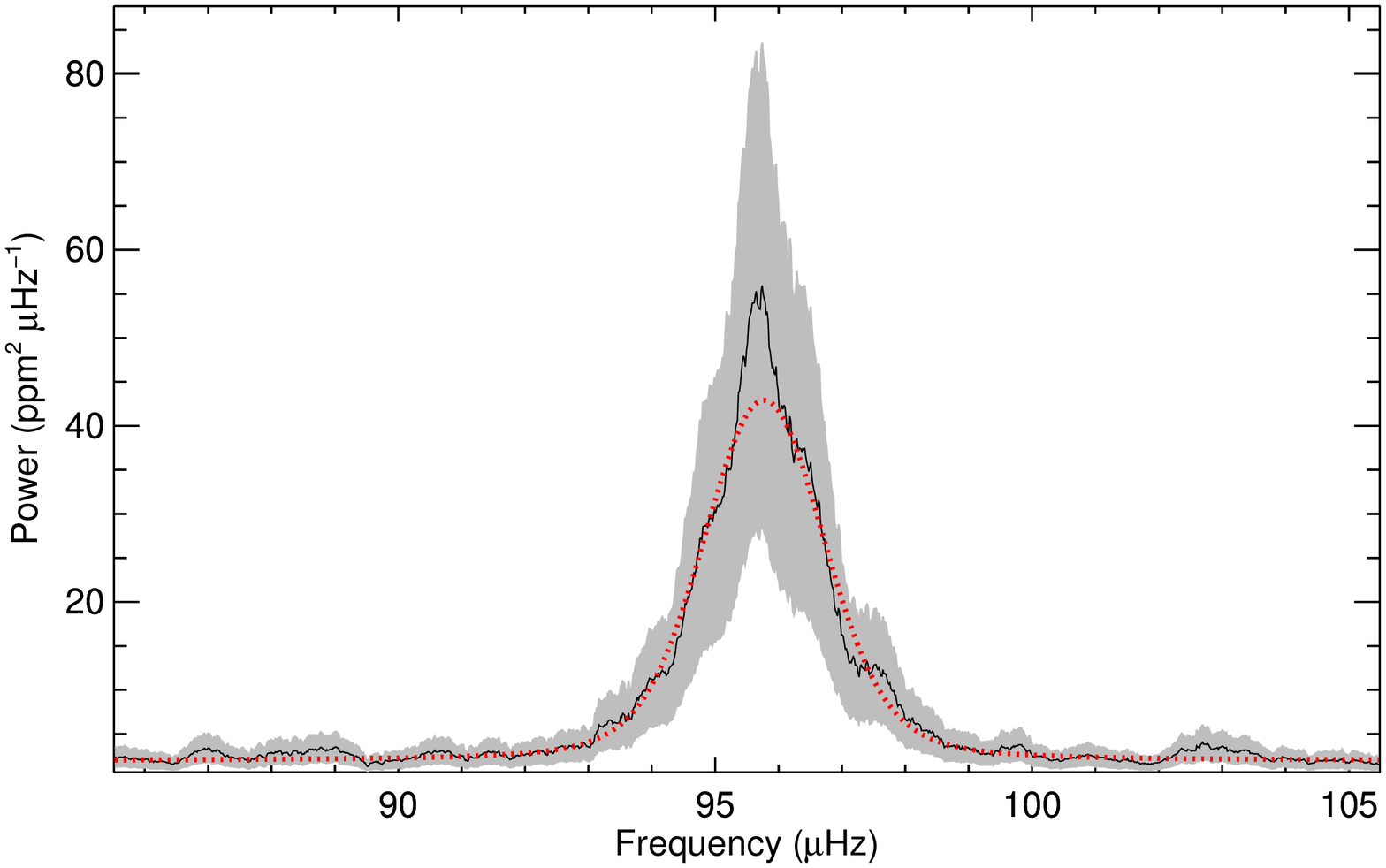}\\
\end{minipage}\hfill
\begin{minipage}[t]{0.33\textwidth}
\includegraphics[angle=270,width=1.05\textwidth]{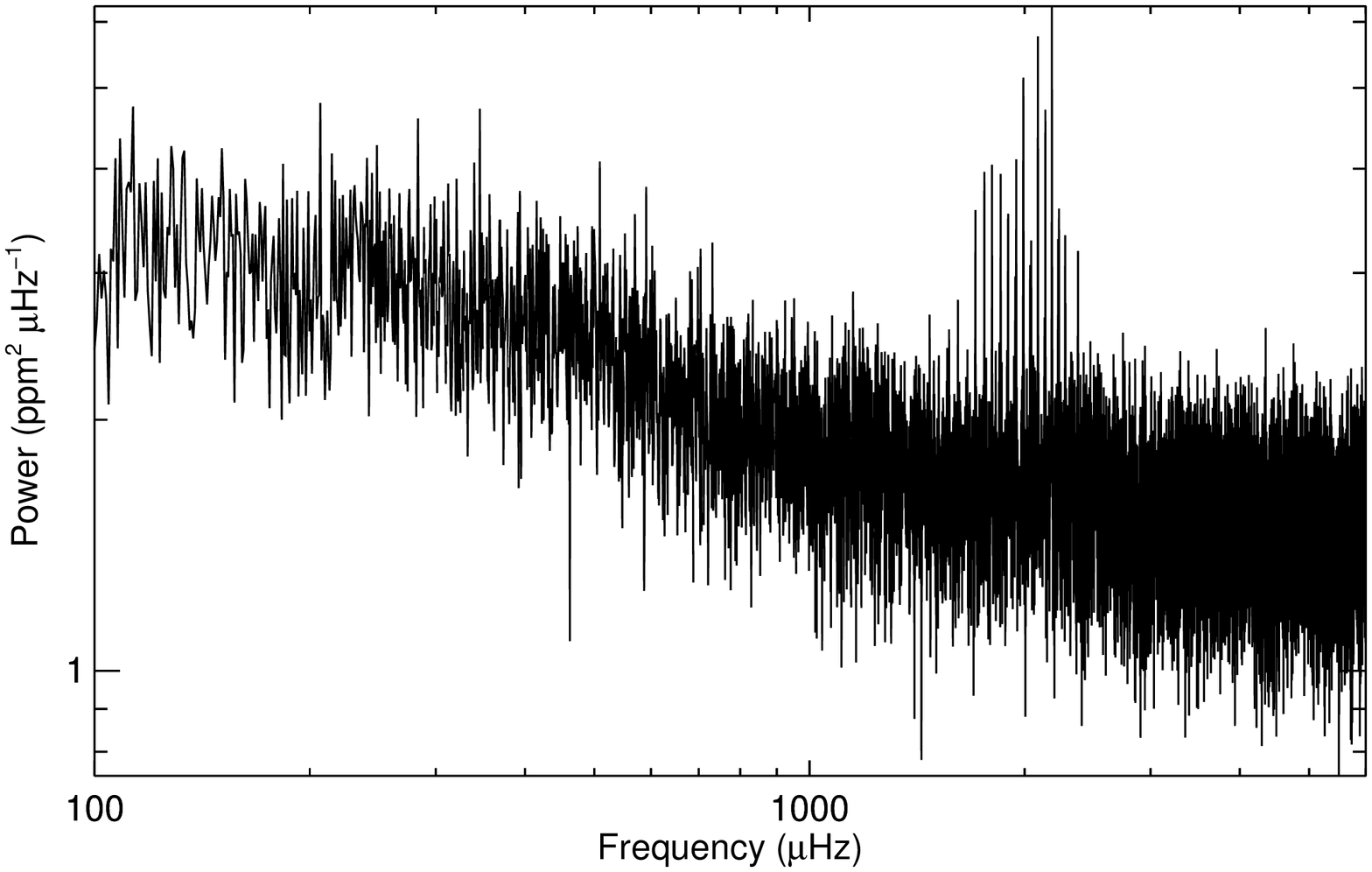}\vspace{-1.em}\\\vspace{-1.em}
\includegraphics[angle=270,width=1.05\textwidth]{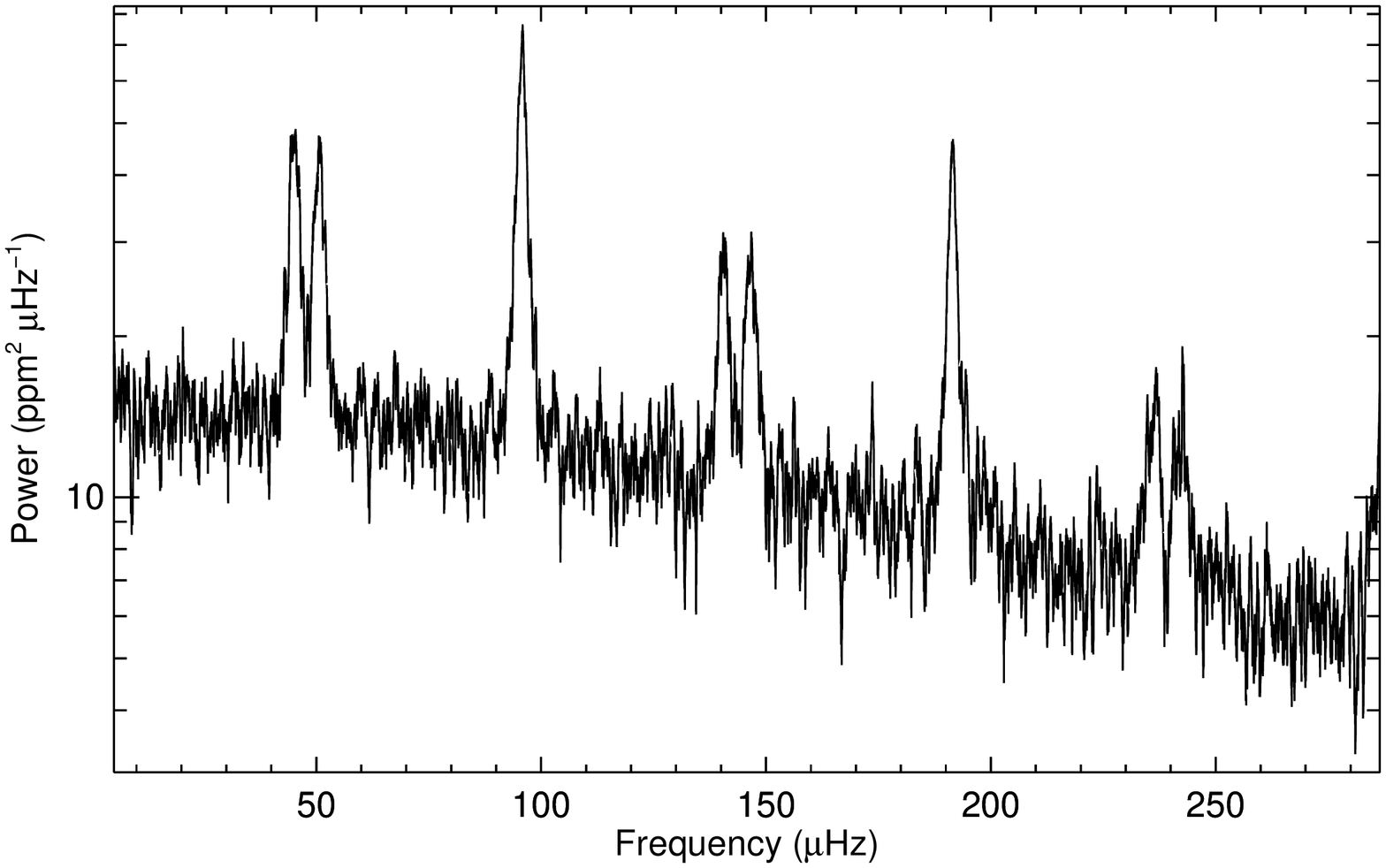}\\
\includegraphics[angle=270,width=1.05\textwidth]{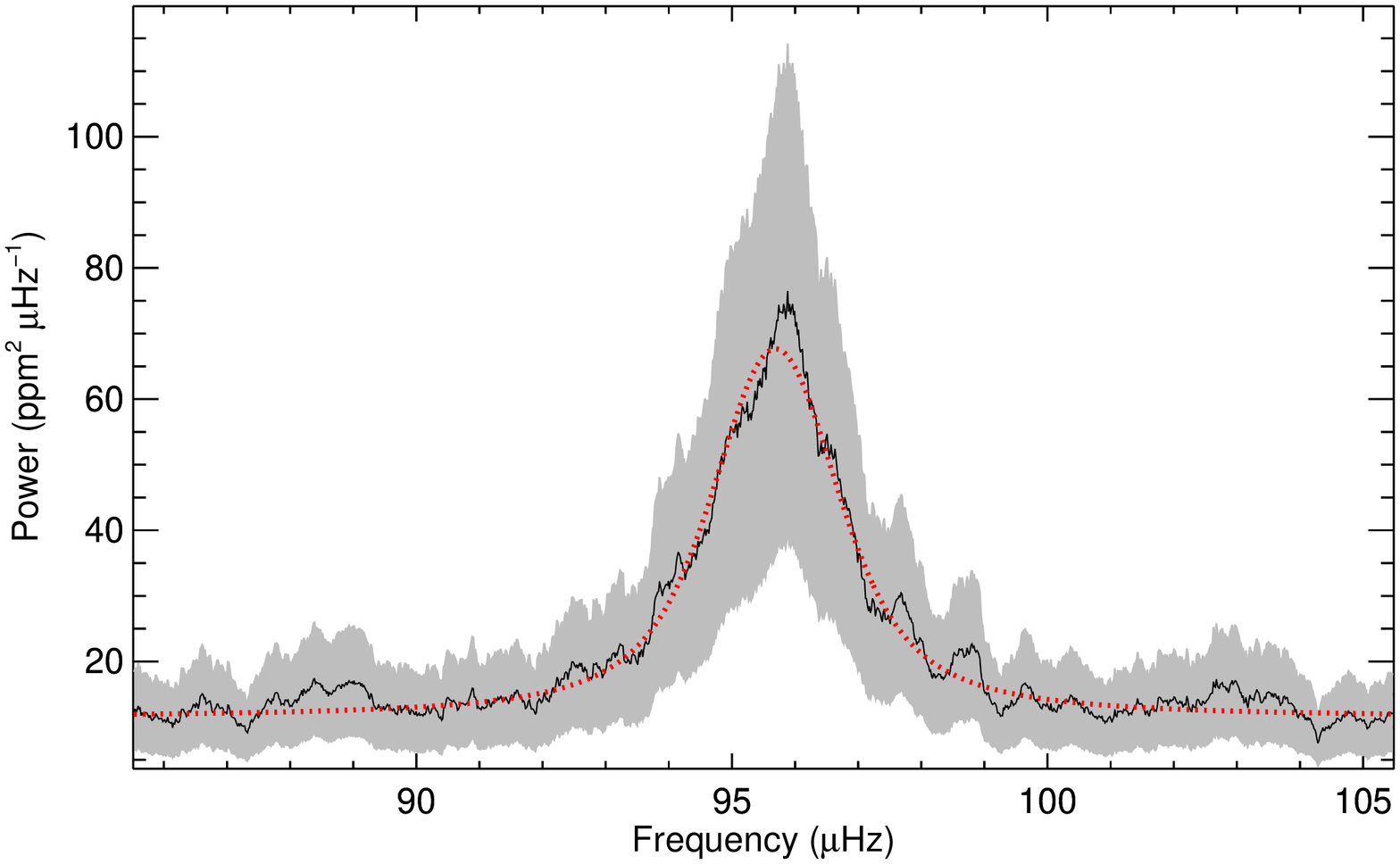}\\
\end{minipage}\hfill
\begin{minipage}[t]{0.33\textwidth}
\includegraphics[angle=270,width=1.05\textwidth]{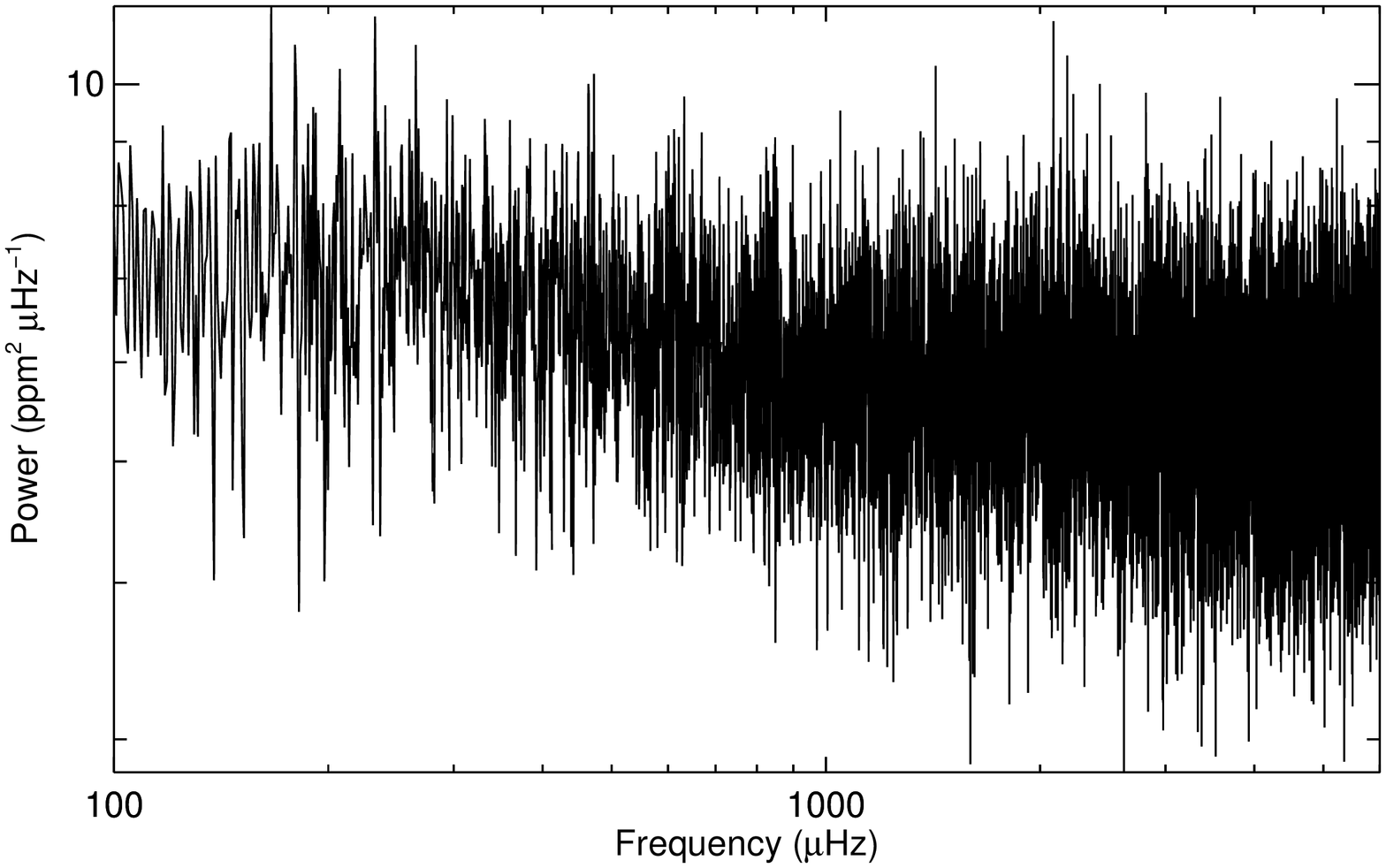}\vspace{-1.em}\\\vspace{-1.em}
\includegraphics[angle=270,width=1.05\textwidth]{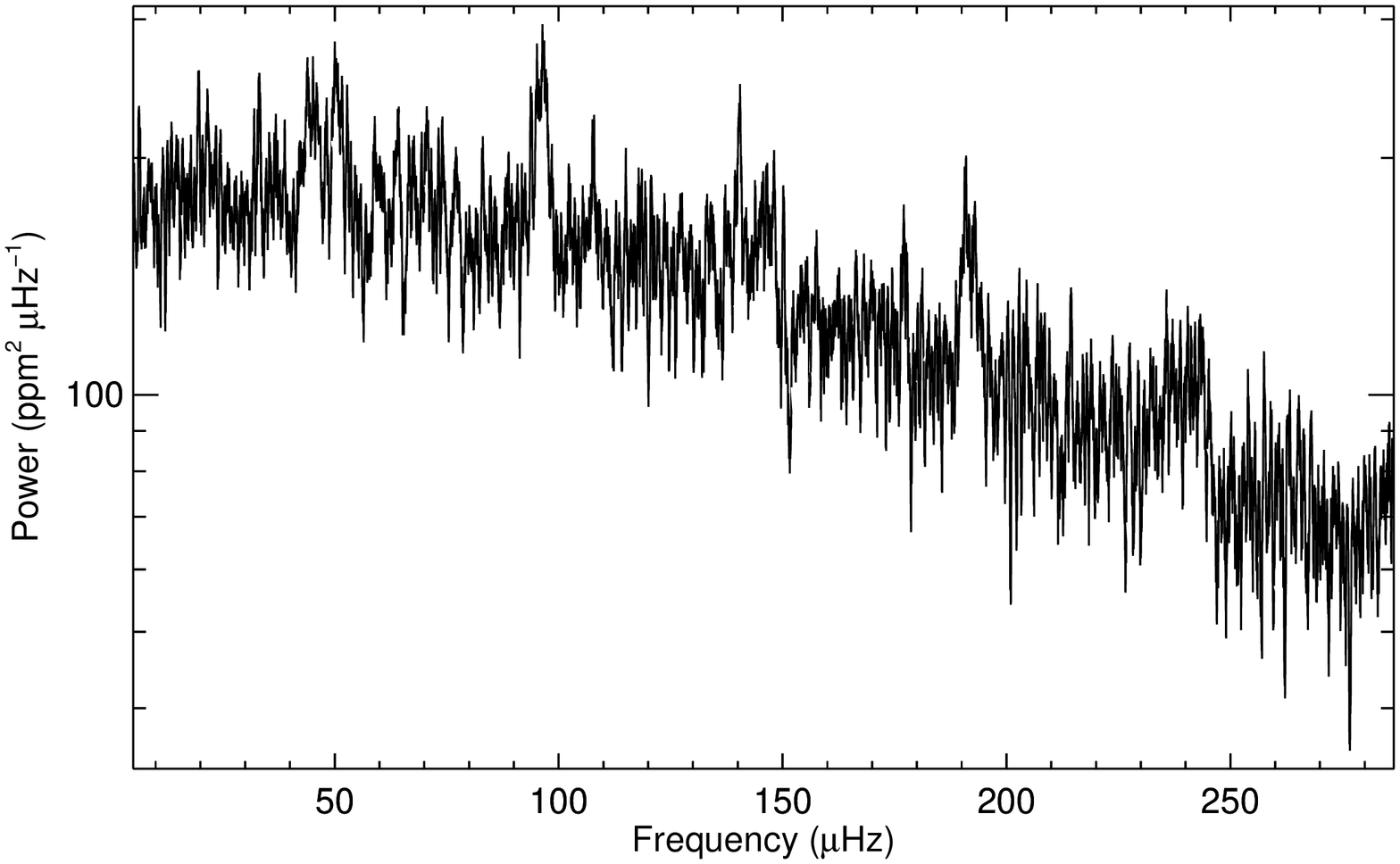}\\
\includegraphics[angle=270,width=1.05\textwidth]{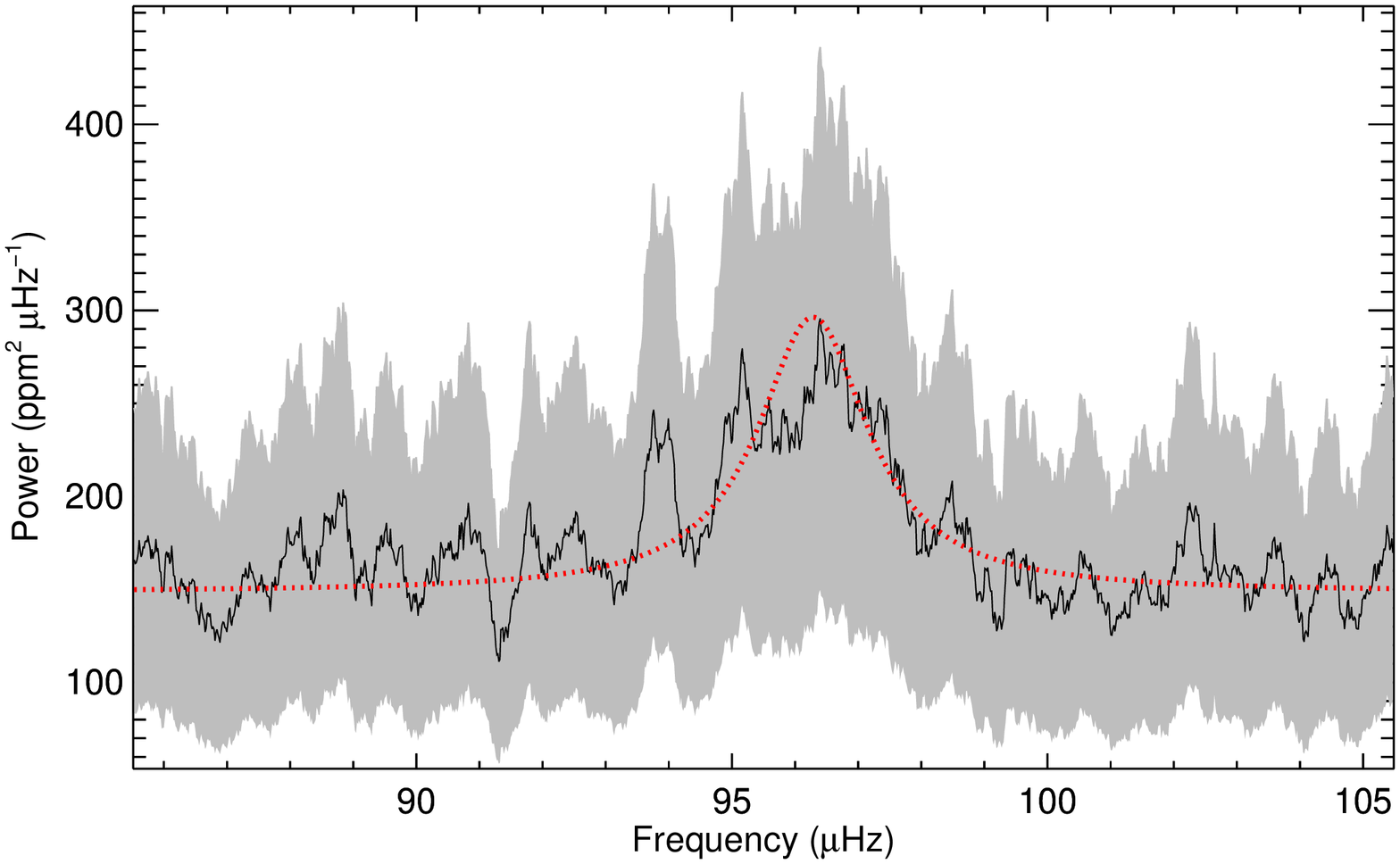}\\
\end{minipage}
\caption{\textbf{Top row:} Double logarithmic plot of the spectra of KIC~5184732 with an artificial S/N of 1, $\frac{1}{4}$, and $\frac{1}{16}$ (from left to right). All spectra are $\unit[0.5]{\mu Hz}$ boxcar smoothed. \textbf{Middle row:} Envelope spectra of the p-mode region between \unit[1.4--2.7]{mHz} of the periodogram shown in the top panel. All envelope spectra are $\unit[0.5]{\mu Hz}$ boxcar smoothed. \textbf{Bottom row:} Fit of Voigt profiles (dashed red line) to the peak of the large frequency separation with $2$-$\sigma$ confidence intervals in grey colour.}
\label{fig:6}
\end{figure*}

In Fig.~\ref{fig:5} a map of the envelope spectrum as a function of central filter frequency is shown which is computed analogously to the procedure described in Sec.~\ref{sec:3.1}. The extended area of large amplitudes in the envelope spectrum at low frequencies and central filter frequencies is because of the higher noise level in stellar data compared to the solar case. Still, large amplitude structures, which indicate regularities in the periodogram, are clearly visible. The large amplitude structure at frequency $\approx\,$\unit[95]{$\mu$Hz}, which corresponds to the large frequency separation, exhibits a small dip at a central filter frequency of $\approx\,$\unit[1600]{$\mu$Hz} followed by a slight slope towards higher frequency difference values for higher values of the central filter frequency. At $\approx\,$\unit[48]{$\mu$Hz} a well-separated double structure can be observed. The separation between these two branches decreases slightly towards higher central filter frequencies, which indicates a decrease in the small separation towards higher frequency. The prime reason for the more distinct splitting of this structure compared to the solar map of the envelope spectrum in Fig.~\ref{fig:3} is the smaller mode linewidth of the p-mode peaks of KIC~5184732 compared to the solar p-modes. 

\subsection{Simulated data, robustness to noise}\label{sec:3.4}
We test the robustness of our approach concerning the signal-to-noise ratio of the time series by generating artificially noisy data sets $x_{\textrm{noisy}}(t)$. For this test, Gaussian white noise with a variance of $a^2\cdot \sigma^2$ is added to the original data,
\begin{align}
x_{\textrm{noisy}}(t) &= x_{\textrm{org}}(t) + a\cdot e(t),\\
 e(t)&\sim N(0,\sigma^2)(t)\, .
\end{align}   
The signal power is defined as the variance $\sigma^2$ of the \textit{Kepler} time series $x_{\textrm{org}}(t)$. The additional noise $e(t)$ has the same variance as the time series, and the scaling factor $a$ specifies the final signal-to-noise ratio,
\begin{align}
\frac{S}{N} = \frac{\textrm{Var}\left(x_{\textrm{org}}\left(t\right)\right)}{\textrm{Var}\left(a\cdot e\left(t\right)\right)}=\frac{\sigma^2}{a^2\cdot\sigma^2} = \frac{1}{a^2}\, .
\end{align}
Even with $\frac{S}{N}=\frac{1}{16}$ the large frequency separation could be accurately obtained.

\begin{table}
\caption{Results for $\Delta\nu$ for KIC~5184732.}           
\label{table:1}   
\centering                                   
\begin{tabular}{c  c  c  c}      
\hline\hline
S/N \quad &  $\Delta\nu$ ($\mu$Hz) \quad &\quad $s_{\Delta\nu}$ ($\mu$Hz) & $\#$overtones\\ 
\hline
Data 	& 95.65   & 0.19 & 3\\
1/1		& 95.65    & 0.19	& 3\\
1/4 	& 95.68   & 0.29	& 2\\
1/16 	& 95.88    & 0.96 & 0\\
\hline                                           
\end{tabular}
\tablefoot{Measured mean large frequency separation $\Delta\nu$ and its error $s_{\Delta\nu}$ for KIC~5184732 for four S/N ratios. The last column indicates the exploited number of overtones in the envelope spectrum.}
\end{table}

In the top row of Fig.~\ref{fig:6}, the double-logarithmic plots of the spectra of the artificially noisy time series for $\frac{S}{N}=1, \frac{1}{4},$ and $\frac{1}{16}$ are shown. These spectra are generated by smoothing the Lomb-Scargle-periodograms with a $\unit[0.5]{\mu Hz}$ boxcar window. The corresponding envelope spectra and the fit for the large frequency separation are presented in the middle and bottom row of Fig.~\ref{fig:6}, respectively. For the computation of the envelope spectra the same frequency range is used as for the original data set. Compared to the spectrum without added noise, shown in Fig.~\ref{fig:4}, the p-mode region is drowned by noise in the spectrum with $S/N=\frac{1}{16}$. However, the frequency separations are still identifiable in the envelope spectrum (middle row, right panel). A fit of the peak, which corresponds to the large frequency separation, gives results comparable to the original data, c.f. Table~\ref{table:1}. In the case of the lowest S/N ratio we can only fit the peak that corresponds to the large frequency separation, while for the original time series three overtones could be exploited as well.

\subsection{Comparison to the autocorrelation of the periodogram}\label{sec:3.5}

\begin{figure*}[ht]
\includegraphics[angle=270,width=0.49\textwidth]{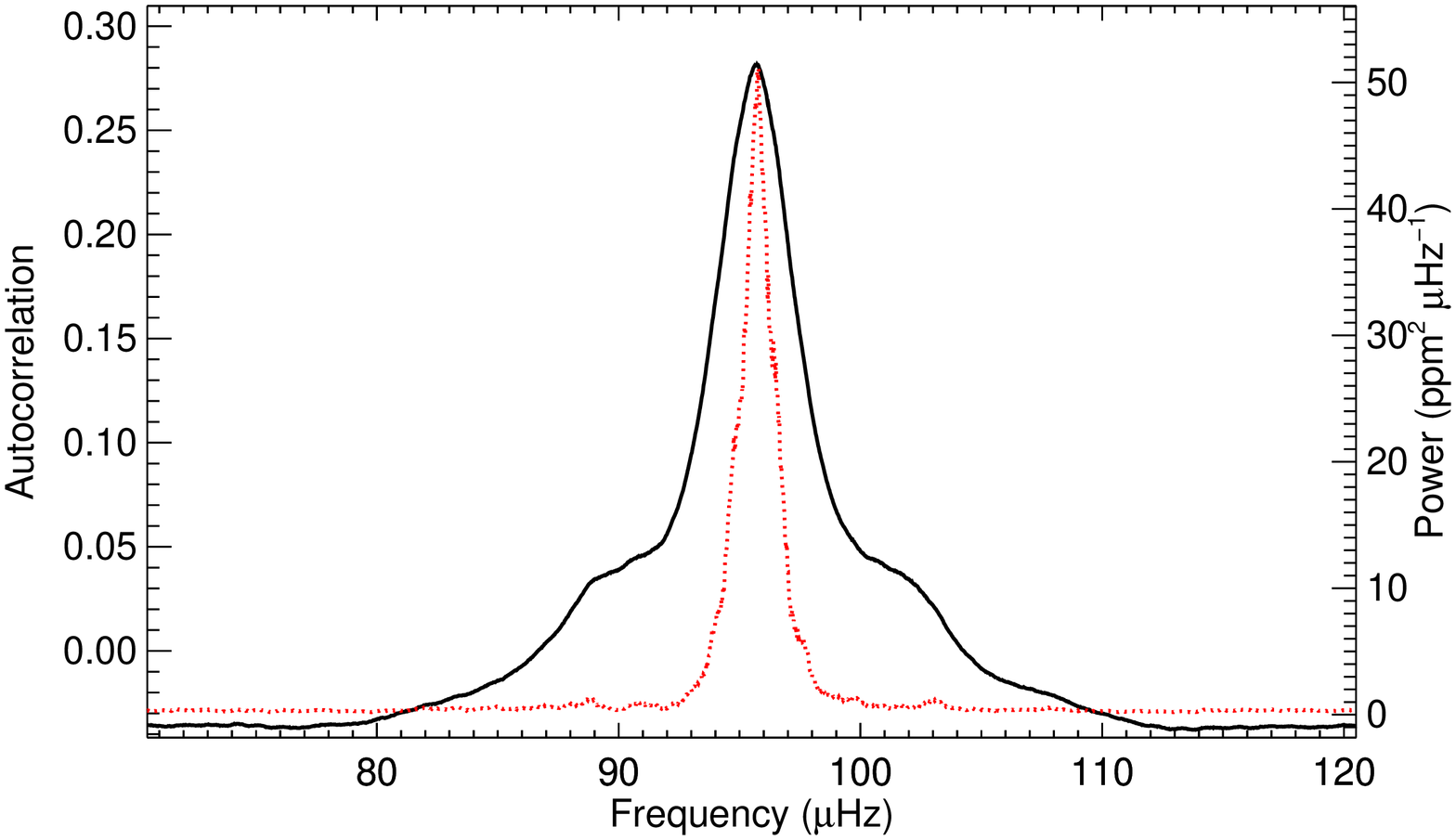}
\includegraphics[angle=270,width=0.49\textwidth]{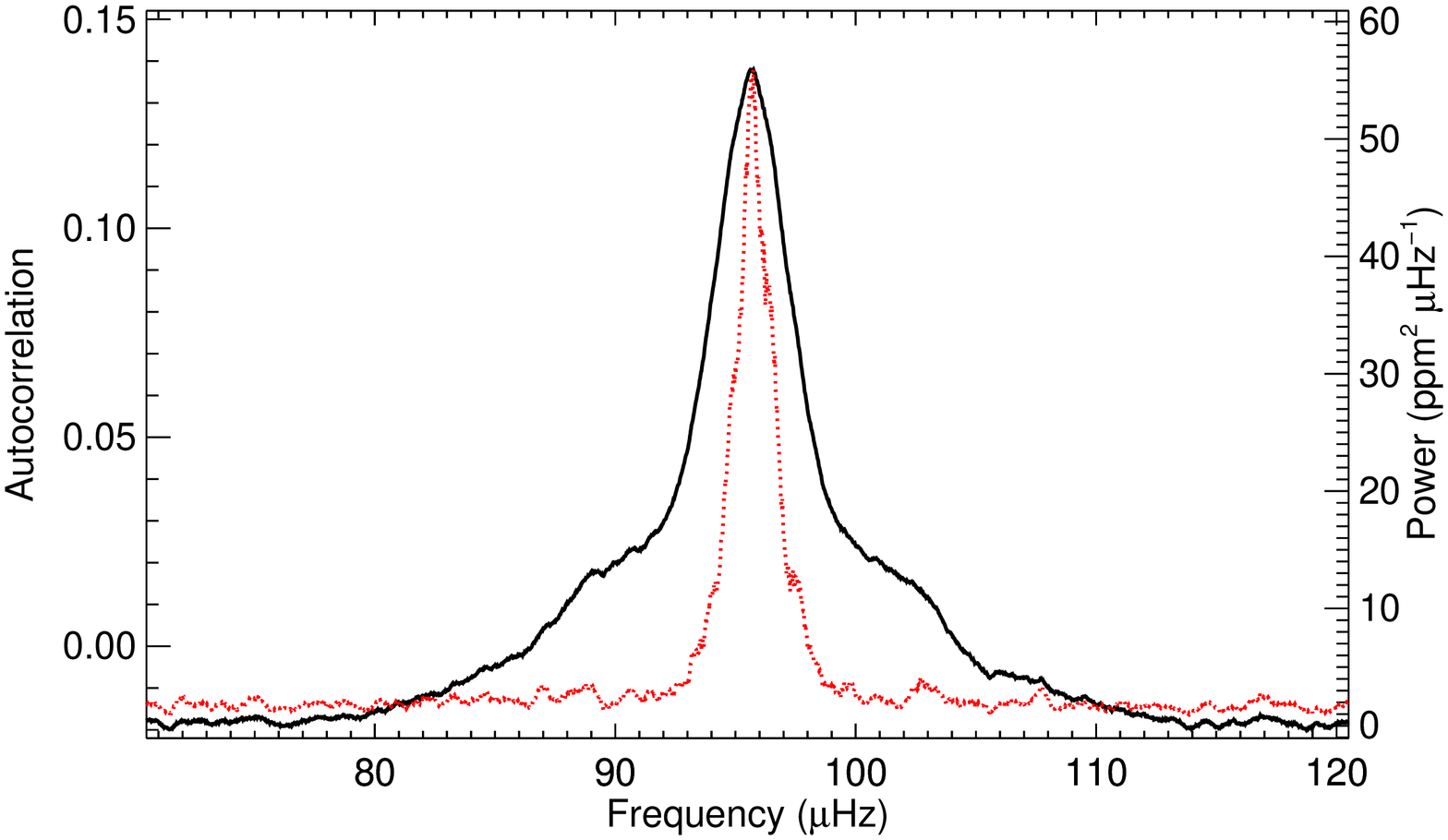}
\includegraphics[angle=270,width=0.49\textwidth]{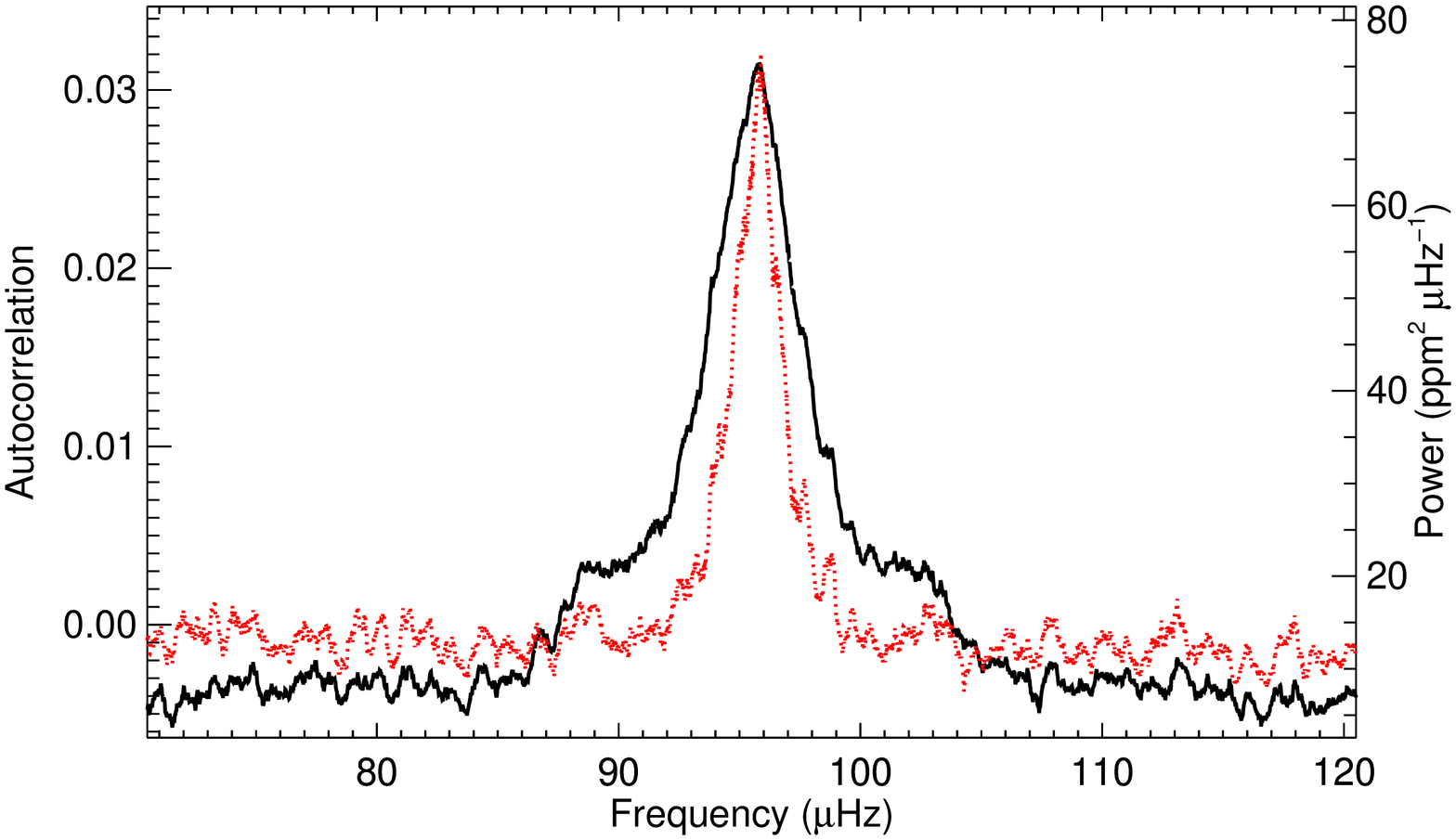}
\includegraphics[angle=270,width=0.49\textwidth]{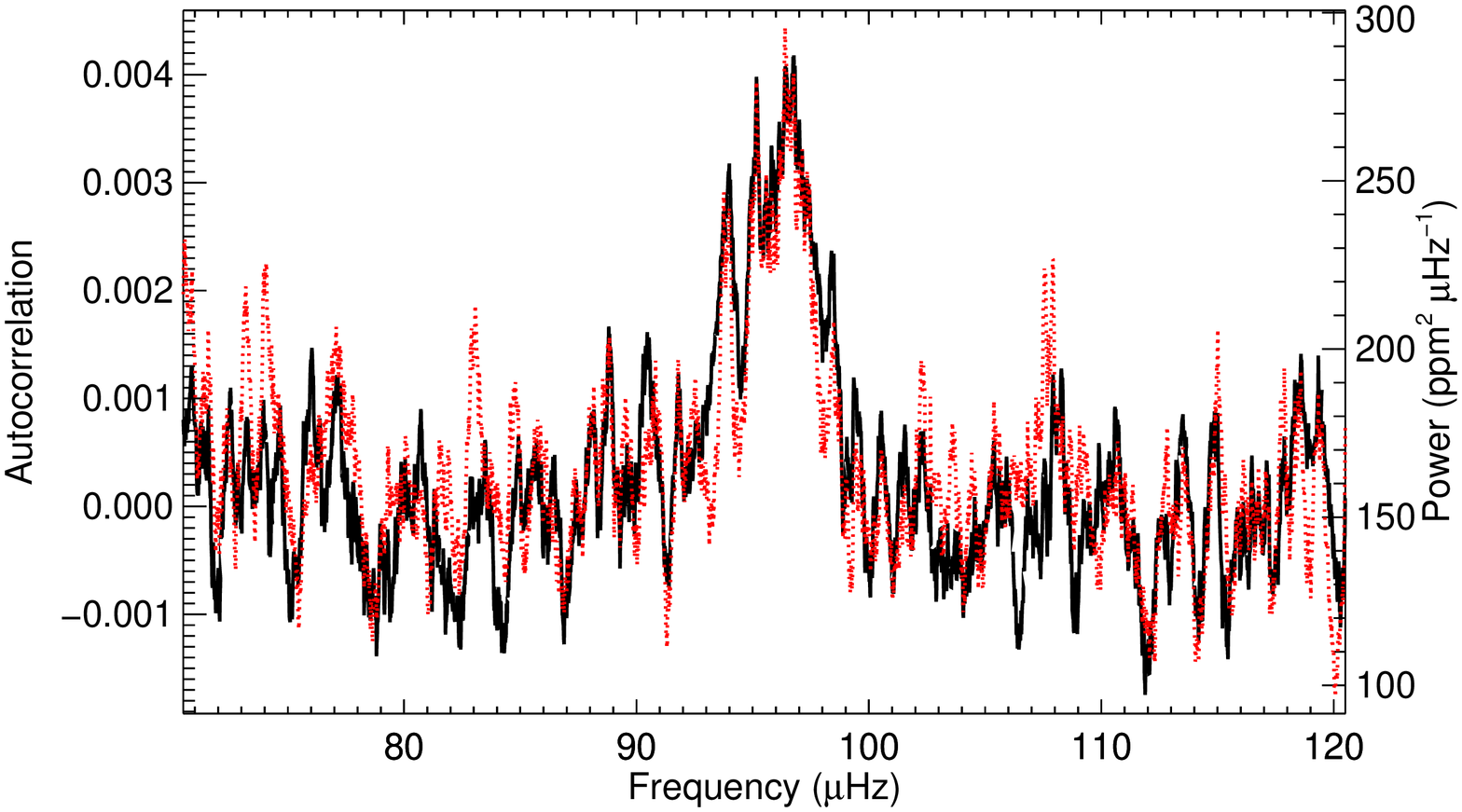}
\caption{Comparison of the ranges around the value of $\Delta\nu$ of the autocorrelation of the periodogram (solid black line) and the corresponding section of the envelope spectrum (dashed red line) for the time series without added noise (top left panel), with $\frac{S}{N}=1$ (top right panel),  with $\frac{S}{N}=\frac{1}{4}$ (bottom left panel), and  with $\frac{S}{N}=\frac{1}{16}$ (bottom right panel). All data is boxcar smoothed over $\unit[0.5]{\mu Hz}$.}
\label{fig:7}
\end{figure*}

We compared the envelope spectrum to an existing method. For this, we computed the autocorrelation of the periodogram of KIC~5184732. The periodogram was filtered in the same way as in the computation of the envelope spectrum. In Fig.~\ref{fig:7} the range around the value of $\Delta\nu$ of the autocorrelation function (ACF) of the periodogram (solid black line) for the time series without added noise (top left panel), with $\frac{S}{N}=1$ (top right panel), with $\frac{S}{N}=\frac{1}{4}$ (bottom left panel), and with $\frac{S}{N}=\frac{1}{16}$ (bottom right panel) are depicted. In direct comparison to this, the corresponding range of the envelope spectrum is plotted as a dashed red line. Both the ACF and the envelope spectra are boxcar smoothed over $\unit[0.5]{\mu Hz}$. A fit of a triple Lorentzian profile to the peak in the ACF of the original time series gave $\Delta\nu = \unit[95.7\pm 2.1]{\mu Hz}$, where the error is given by the HWHM of the central Lorentzian. In contrast to this the HWHM of the peak in the envelope spectrum is only $\unit[1.1]{\mu Hz}$.

For the artificially noisy time series with $\frac{S}{N}=1$ and with $\frac{S}{N}=\frac{1}{4}$, the peak in the envelope spectrum is narrower than in the ACF. For the artificially low $\frac{S}{N}=\frac{1}{16}$ only, the HWHM of both peaks become equally large (bottom right panel of Fig.~\ref{fig:7}).

Consequently, the error on the large frequency separation extracted from the envelope spectrum is smaller or equal to the corresponding value extracted from the ACF.

\section{Discussion and conclusion}\label{sec:4}
The envelope spectrum allows a reliable detection of regularities in the periodograms of solar and stellar oscillation time series. For solar-like oscillators the mean small and large frequency separations are the prime candidates for these kinds of regularities in the periodogram. 

Applied to data measured with the GOLF instrument, we obtained a mean solar large frequency separation of $\Delta\nu_{\odot} = \unit[134.92\pm 0.06]{\mu Hz}$. If overtones are present in the envelope spectrum at multiples of the fundamental frequency, these can be used to decrease the error on the extracted value on the frequency separation, since the overtones are independent of each other. 

For the Kepler star KIC~5184732, we measured the mean large frequency separation to $\Delta\nu = \unit[95.7\pm 0.2]{\mu Hz}$. This value is obtained with good agreement even if the S/N of the data is artificially lowered. This result shows the advantage of the presented method regarding stellar time series with low S/N. Even if an excess of power due to p-modes is not visible by eye in the periodogram, the envelope spectrum can accurately measure the hidden large frequency separation. The robustness of the method to noise in the data makes it a viable approach to obtain p-mode parameters from stars with a low S/N of the p-modes. In comparison to the autocorrelation of the periodogram, we find that the error on the large frequency separation extracted from the envelope spectrum is smaller or equal to the corresponding value extracted from the ACF. This is an indication that the envelope spectrum performs somewhat better than the ACF in the extraction of the large frequency separation.

We also showed that by incrementally shifting the frequency filter through the periodogram, regularities in the periodogram can be visualised. This map of regularities and their frequency dependence can be created without the choice of a critical parameter such as the large frequency separation for echelle diagrams. It is particularly useful for a first visual estimation of the frequency range of p-modes and of $\Delta\nu$.

\begin{acknowledgements}
The research leading to these results received funding from the European Research Council under the European Union’s Seventh Framework Program (FP/2007-2013)/ERC Grant Agreement no. 307117. 
\end{acknowledgements}
\bibliographystyle{aa}
\bibliography{references} 

\begin{thebibliography}{18}
\expandafter\ifx\csname natexlab\endcsname\relax\def\natexlab#1{#1}\fi

\bibitem[{{Christensen-Dalsgaard} {et~al.}(1996){Christensen-Dalsgaard},
  {Dappen}, {Ajukov}, {Anderson}, {Antia}, {Basu}, {Baturin}, {Berthomieu},
  {Chaboyer}, {Chitre}, {Cox}, {Demarque}, {Donatowicz}, {Dziembowski},
  {Gabriel}, {Gough}, {Guenther}, {Guzik}, {Harvey}, {Hill}, {Houdek},
  {Iglesias}, {Kosovichev}, {Leibacher}, {Morel}, {Proffitt}, {Provost},
  {Reiter}, {Rhodes}, {Rogers}, {Roxburgh}, {Thompson}, \&
  {Ulrich}}]{1996Sci...272.1286C}
{Christensen-Dalsgaard}, J., {Dappen}, W., {Ajukov}, S.~V., {et~al.} 1996,
  Science, 272, 1286

\bibitem[{{Garc{\'{\i}}a} {et~al.}(2005){Garc{\'{\i}}a}, {Turck-Chi{\`e}ze},
  {Boumier}, {Robillot}, {Bertello}, {Charra}, {Dzitko}, {Gabriel},
  {Jim{\'e}nez-Reyes}, {Pall{\'e}}, {Renaud}, {Roca Cort{\'e}s}, \&
  {Ulrich}}]{2005A&A...442..385G}
{Garc{\'{\i}}a}, R.~A., {Turck-Chi{\`e}ze}, S., {Boumier}, P., {et~al.} 2005,
  \aap, 442, 385

\bibitem[{Harris(1978)}]{Harris}
Harris, F. 1978, Proceedings of the IEEE, 66, 51

\bibitem[{{Hekker} {et~al.}(2011){Hekker}, {Elsworth}, {De Ridder}, {Mosser},
  {Garc{\'{\i}}a}, {Kallinger}, {Mathur}, {Huber}, {Buzasi}, {Preston}, {Hale},
  {Ballot}, {Chaplin}, {R{\'e}gulo}, {Bedding}, {Stello}, {Borucki}, {Koch},
  {Jenkins}, {Allen}, {Gilliland}, {Kjeldsen}, \&
  {Christensen-Dalsgaard}}]{2011A&A...525A.131H}
{Hekker}, S., {Elsworth}, Y., {De Ridder}, J., {et~al.} 2011, \aap, 525, A131

\bibitem[{{Houdek} {et~al.}(1999){Houdek}, {Balmforth},
  {Christensen-Dalsgaard}, \& {Gough}}]{1999A&A...351..582H}
{Houdek}, G., {Balmforth}, N.~J., {Christensen-Dalsgaard}, J., \& {Gough},
  D.~O. 1999, \aap, 351, 582

\bibitem[{{Huber} {et~al.}(2009){Huber}, {Stello}, {Bedding}, {Chaplin},
  {Arentoft}, {Quirion}, \& {Kjeldsen}}]{2009CoAst.160...74H}
{Huber}, D., {Stello}, D., {Bedding}, T.~R., {et~al.} 2009, Communications in
  Asteroseismology, 160, 74

\bibitem[{{Kallinger} {et~al.}(2010){Kallinger}, {Mosser}, {Hekker}, {Huber},
  {Stello}, {Mathur}, {Basu}, {Bedding}, {Chaplin}, {De Ridder}, {Elsworth},
  {Frandsen}, {Garc{\'{\i}}a}, {Gruberbauer}, {Matthews}, {Borucki}, {Bruntt},
  {Christensen-Dalsgaard}, {Gilliland}, {Kjeldsen}, \&
  {Koch}}]{2010A&A...522A...1K}
{Kallinger}, T., {Mosser}, B., {Hekker}, S., {et~al.} 2010, \aap, 522, A1

\bibitem[{{Komm} {et~al.}(2000){Komm}, {Howe}, \& {Hill}}]{2000ApJ...543..472K}
{Komm}, R.~W., {Howe}, R., \& {Hill}, F. 2000, \apj, 543, 472

\bibitem[{{Lomb}(1976)}]{1976Ap&SS..39..447L}
{Lomb}, N.~R. 1976, \apss, 39, 447

\bibitem[{{Mathur} {et~al.}(2010){Mathur}, {Garc{\'{\i}}a}, {R{\'e}gulo},
  {Creevey}, {Ballot}, {Salabert}, {Arentoft}, {Quirion}, {Chaplin}, \&
  {Kjeldsen}}]{2010A&A...511A..46M}
{Mathur}, S., {Garc{\'{\i}}a}, R.~A., {R{\'e}gulo}, C., {et~al.} 2010, \aap,
  511, A46

\bibitem[{{Mathur} {et~al.}(2012){Mathur}, {Metcalfe}, {Woitaszek}, {Bruntt},
  {Verner}, {Christensen-Dalsgaard}, {Creevey}, {Do{\v g}an}, {Basu}, {Karoff},
  {Stello}, {Appourchaux}, {Campante}, {Chaplin}, {Garc{\'{\i}}a}, {Bedding},
  {Benomar}, {Bonanno}, {Deheuvels}, {Elsworth}, {Gaulme}, {Guzik}, {Handberg},
  {Hekker}, {Herzberg}, {Monteiro}, {Piau}, {Quirion}, {R{\'e}gulo}, {Roth},
  {Salabert}, {Serenelli}, {Thompson}, {Trampedach}, {White}, {Ballot},
  {Brand{\~a}o}, {Molenda-{\.Z}akowicz}, {Kjeldsen}, {Twicken}, {Uddin}, \&
  {Wohler}}]{2012ApJ...749..152M}
{Mathur}, S., {Metcalfe}, T.~S., {Woitaszek}, M., {et~al.} 2012, \apj, 749, 152

\bibitem[{{Mosser} \& {Appourchaux}(2009)}]{2009A&A...508..877M}
{Mosser}, B. \& {Appourchaux}, T. 2009, \aap, 508, 877

\bibitem[{{Mosser} {et~al.}(2013){Mosser}, {Michel}, {Belkacem}, {Goupil},
  {Baglin}, {Barban}, {Provost}, {Samadi}, {Auvergne}, \&
  {Catala}}]{2013A&A...550A.126M}
{Mosser}, B., {Michel}, E., {Belkacem}, K., {et~al.} 2013, \aap, 550, A126

\bibitem[{Priestley(1982)}]{Priestley}
Priestley, B. 1982, Spectral analysis and time series, Probability and
  mathematical statistics No. Bd. 1-2 (Academic Press)

\bibitem[{{Scargle}(1982)}]{1982ApJ...263..835S}
{Scargle}, J.~D. 1982, \apj, 263, 835

\bibitem[{{Tassoul}(1980)}]{1980ApJS...43..469T}
{Tassoul}, M. 1980, \apjs, 43, 469

\bibitem[{{Verner} {et~al.}(2011){Verner}, {Elsworth}, {Chaplin}, {Campante},
  {Corsaro}, {Gaulme}, {Hekker}, {Huber}, {Karoff}, {Mathur}, {Mosser},
  {Appourchaux}, {Ballot}, {Bedding}, {Bonanno}, {Broomhall}, {Garc{\'{\i}}a},
  {Handberg}, {New}, {Stello}, {R{\'e}gulo}, {Roxburgh}, {Salabert}, {White},
  {Caldwell}, {Christiansen}, \& {Fanelli}}]{2011MNRAS.415.3539V}
{Verner}, G.~A., {Elsworth}, Y., {Chaplin}, W.~J., {et~al.} 2011, \mnras, 415,
  3539

\bibitem[{{Verner} \& {Roxburgh}(2011)}]{2011arXiv1104.0631V}
{Verner}, G.~A. \& {Roxburgh}, I.~W. 2011, ArXiv e-prints
  [\eprint[arXiv]{1104.0631}]

\end{thebibliography}

\end{document}